\documentclass[acmlarge,screen]{acmart}
\usepackage{bbding}
\usepackage{makecell}
\usepackage{hyperref}
\usepackage{subfigure}
\usepackage{algorithm}
\usepackage{algpseudocode}
\usepackage{soul, color, xcolor}

\AtBeginDocument{%
  }

\setcopyright{acmlicensed}
\acmJournal{IMWUT}
\acmYear{2025} 
\acmVolume{9} 
\acmNumber{3} 
\acmArticle{81} 
\acmMonth{9}
\acmDOI{10.1145/3749504}

\begin{document}

\title{mmPencil: Toward Writing-Style-Independent In-Air Handwriting Recognition via mmWave Radar and Large Vision-Language Model}
\renewcommand{\shorttitle}{mmPencil: Toward Writing-Style-Independent In-Air Handwriting Recognition via mmWave Radar and VLMs}


\author{Yifan Guo}
\orcid{0009-0009-0284-8965}
\affiliation{%
 \institution{Northwestern Polytechnical University}
 \city{Xi'an}
 \state{Shaanxi}
 \country{China}
 \postcode{710129}
}
\email{guoyf@mail.nwpu.edu.cn}

\author{Zhu Wang}
\authornote{Corresponding author}
\orcid{0000-0003-2368-8947}
\affiliation{%
 \institution{Northwestern Polytechnical University}
 \city{Xi'an}
 \state{Shaanxi}
 \country{China}
 \postcode{710129}
}
\email{wangzhu@nwpu.edu.cn}

\author{Qian Qin}
\orcid{0009-0000-0420-2134}
\affiliation{%
 \institution{Northwestern Polytechnical University}
 \city{Xi'an}
 \state{Shaanxi}
 \country{China}
 \postcode{710129}
}
\email{qianqin@mail.nwpu.edu.cn}

\author{Yangqian Lei}
\orcid{0009-0003-8171-0230}
\affiliation{%
 \institution{Northwestern Polytechnical University}
 \city{Xi'an}
 \state{Shaanxi}
 \country{China}
 \postcode{710129}
}
\email{leiyangqian@mail.nwpu.edu.cn}

\author{Qiwen Gan}
\orcid{0009-0007-4404-4303}
\affiliation{%
 \institution{Northwestern Polytechnical University}
 \city{Xi'an}
 \state{Shaanxi}
 \country{China}
 \postcode{710129}
}
\email{ganqw@mail.nwpu.edu.cn}

\author{Zhuo Sun}
\orcid{0000-0003-3115-278X}
\affiliation{%
 \institution{Northwestern Polytechnical University}
 \city{Xi'an}
 \state{Shaanxi}
 \country{China}
 \postcode{710129}
}
\email{zsun@nwpu.edu.cn}

\author{Chao Chen}
\orcid{0000-0003-2094-9734}
\affiliation{%
 \institution{Chongqing University}
 \state{Chongqing}
 \country{China}
}
\email{cschaochen@cqu.edu.cn}

\author{Bin Guo}
\orcid{0000-0001-6097-2467}
\affiliation{%
 \institution{Northwestern Polytechnical University}
 \city{Xi'an}
 \state{Shaanxi}
 \country{China}
 \postcode{710129}
}
\email{guob@nwpu.edu.cn}

\author{Zhiwen Yu}
\orcid{0000-0002-9905-3238}
\affiliation{%
 \institution{Northwestern Polytechnical University}
 \city{Xi'an}
 \state{Shaanxi}
 \country{China}
 \postcode{710129}
}
\affiliation{%
 \institution{Harbin Engineering University}
 \city{Harbin}
 \state{Heilongjiang}
 \country{China}
 \postcode{150001}
}
\email{zhiwenyu@nwpu.edu.cn}

\renewcommand{\shortauthors}{Guo et al.}

\begin{abstract}
With the rapid advancement of wireless technologies, in-air handwriting recognition based on radio frequency (RF) signals emerges as a promising solution for human-computer interaction. However, existing methods often constrain writing gestures to predefined planes, impose strict requirements on writing order and direction, and limit recognition tasks to a small set of digits or letters. To overcome these limitations, we propose $m^2VLMs$, a modality mapping architecture, which serves as a bridge between the millimeter-wave (mmWave) radar sensing technologies and large vision-language models (VLMs). Based on the proposed architecture, a three-dimensional (3D) in-air handwriting word recognition system named \textit{mmPencil} is developed. Specifically, we design a multi-stage spatial trajectory reconstruction algorithm that extracts frequency-domain features to identify handwriting regions and achieve high-precision reconstruction of 3D word trajectories. Furthermore, we introduce a novel spatial-to-visual mapping algorithm, which bridges the gap between spatial trajectory information captured by mmWave radar and vision-language representations, providing a foundation for cross-modality understanding in the large vision-language model. As a result, \textit{mmPencil} tackles the limitations of current solutions that are heavily reliant on handwriting styles and environmental factors, expanding RF-based in-air handwriting recognition to more complex word-level scenarios. 

We collect and release a 3D mmWave handwriting dataset comprising 200 distinct words (ranging from 2 to 9 letters), contributions from 12 users, and 22 different writing scenarios, totaling 7,664 samples with an overall size of 31.66 GB. Extensive experiments demonstrate that \textit{mmPencil} achieves accurate and robust word recognition in real-world scenarios, remaining unaffected by variations in word categories, length, as well as writing position, range, angle, speed, size, direction, and user movement. Specifically, for 4 seen users, \textit{mmPencil} achieves a recognition accuracy of approximately 97.60\% across 200 word classes. Moreover, benefiting from the generalization capability of VLMs, the zero-shot recognition accuracy for 4 unseen users can also reach 92.50\% across 50 word classes using training data from only 8 users, which outperforming state-of-the-art baselines.

\end{abstract}




\maketitle

\section{Introduction} \label{introduction}
With the rapid development of information technology, human-computer interaction (HCI) has become increasingly important in fields such as smart devices \cite{ng2017smart}, industrial internet of things (IoT) \cite{mucha2018industrial}, and assistive technologies \cite{reyes2022demonstrating}. Traditional interaction methods rely on input devices like touchscreens, mice, and keyboards. However, these methods perform poorly on devices with limited screen sizes (e.g., smartwatches) or without dedicated input interfaces, and they are not always user-friendly for specific groups such as the elderly and people with disabilities. As a result, innovative interaction technologies, such as in-air writing recognition\cite{arsalan2019character}, voice input \cite{jacob1996human,cohen1995role}, and visual gesture recognition \cite{pavlovic1997visual}, have garnered growing attention. However, voice and visual-based methods can leak sensitive personal information, such as voice tone, conversation content, or facial features, posing privacy risks. In contrast, in-air writing recognition based on radio frequency (RF) signals offers the advantages of non-contact and privacy protection, which allows the user to write in the air across various scenarios. It overcomes many limitations of traditional approaches and has become an ideal choice for achieving natural human-computer interaction.

Unlike common non-interactive sensing tasks such as vital signs monitoring \cite{song2024finersense,wang2024size,ren2024characterizing,zhang2023understanding}, material detection \cite{wang2024liqdetector,wang2024grainsense,wang2024wi2dmeasure,chen2024metabioliq}, and human activity recognition \cite{chang2024robust,wang2024xrf55,xu2019acousticid}, in-air writing recognition is a more fine-grained sensing task. By analyzing reflected signals from the target, it deciphers handwriting content and provides richer semantic information. In-air writing recognition has been extensively studied in the literature. Early approaches utilized acoustic devices \cite{du2018wordrecorder,yin2020learning,wang2023audiowrite,ding2019assv} to extract distinctive acoustic patterns related to writing trajectories, enabling the classification of letters or words. However, this line of methods is highly susceptible to environmental noise. Solutions based on RF signals commonly utilize Wi-Fi channel state information (CSI) \cite{wu2020fingerdraw,han2021towards,zhang2022handgest,regani2022gwrite} to detect fine finger movements by leveraging detailed subcarrier information. However, Wi-Fi signals are highly affected by severe multipath effects and have limited sensing resolution. Similarly, RFID-based approaches \cite{wang2018multi} require a large number of tag arrays, which makes them difficult to deploy in real-world scenarios. In contrast, mmWave radar has recently gained widespread attention for its rich sensing capabilities, including range, angle, and velocity, along with its high resolution. Beyond in-air writing recognition, it has demonstrated outstanding performance in applications such as heartbeat monitoring \cite{zhao2024mmarrhythmia,chang2024mmecare,zhao2024airecg}, vibration detection \cite{yang2023multi,jiang2020mmvib}, and facial recognition \cite{xu2022mask,xie2023mm3dface,zhang2023mmfer}.

To extract handwriting trajectory information, existing radar-based studies can be roughly divided into two categories, i.e., the model-driven method and the data-driven method. Model-driven methods construct physical models \cite{lin2024airwrite,regani2021mmwrite} between the signals and the handwriting trajectory to perform handwriting tracking and recognition. This line of studies, while eliminating the need for pre-training on collected data, struggle to handle characters written in non-predefined orders and are heavily reliant on specific scenarios. On the contrary, data-driven methods transform RF signals into trajectories or spectrograms and build classifiers based on long short-term memory (LSTM) or convolution neural network (CNN) \cite{arsalan2019character,wei2022indexpen,chen2023mmhtsr}. Nevertheless, systems based on data-driven methods rely heavily on large-scale and high-quality training datasets. However, RF signals are highly susceptible to interference from factors such as writing position, direction, and individual writing habits. The scarcity and low quality of the RF sensing data, along with the diversity of handwriting content, greatly limit the development of RF-based in-air handwriting recognition.

Recently, foundation models (FMs), represented by large language models (LLMs) and large vision-language models (VLMs), have demonstrated unique capabilities in processing sensing data from the physical world, offering new opportunities for in-air writing recognition. For instance, the Tent system \cite{zhou2023tent} integrates IoT sensor signals with text embeddings to enable zero-shot human activity recognition, while Penetrative AI \cite{xu2024penetrative} connects LLMs with the physical world, empowering them to process and interpret sensory data. Similarly, the FM-Fi system \cite{weng2024large} introduces an innovative cross-modal system to transfer knowledge from vision-FMs, which facilitates human activity recognition based on RF signals. However, the powerful capabilities emerging from FMs such as VLMs have not yet been applied to in-air writing recognition.

Although there has been work on enhancing wireless sensing using LLMs or VLMs, achieving writing-style-independent in-air handwriting recognition via the mmWave radar and the large vision-language model is still a non-trivial issue, due to the following three major challenges. First, realistic in-air writing usually occurs in 3D space, making it difficult to accurately reconstruct writing trajectories from composite RF signals, even with a mmWave radar featuring a 4GHz bandwidth and an antenna array. Second, the reflected signals of handwriting movements depend on not only the word being written, but also factors such as writing position, direction, and order, resulting in significant variability even within samples of the same word. Third, the RF data differs significantly from the text and image data, which inherently possess abundant semantic information. This disparity restricts the efficacy of VLMs when directly handling RF signals.

To address the aforementioned challenges, we develop a system named \textit{mmPencil}, which is a pioneering effort in harnessing the powerful generalization capability of VLMs to achieve writing-style-independent recognition of in-air handwriting words. Specifically, a modality mapping architecture is proposed to bridge the mmWave data and VLMs, which consists of four components: data collection, trajectory reconstruction, data mapping, and instruction fine-tuning. First, we collect and release a 3D mmWave in-air handwriting word dataset. Afterwards, an environment-robust multi-stage spatial trajectory reconstruction algorithm is designed, which effectively mitigates the interference caused by variations in handwriting positions. In particular, by exploring the range-azimuth and range-elevation profiles refined through spectral refinement, we accurately capture the spatial trajectory of handwriting movements at each moment. Outlier removal is performed on the obtained trajectory sequence, followed by smoothing using a 3D Kalman filter. Next, we design a spatial-to-visual mapping algorithm to achieve handwriting recognition independent of writing styles, such as direction and order. Through direction-corrected Principal Component Analysis (PCA), we map the 3D trajectories into 2D images while preserving the shape features of the handwriting words. This mapping algorithm bridges the gap between wireless signals and visual images, enabling VLMs to intuitively interpret and process RF signals. Finally, we put forward an instruction fine-tuning approach to improve handwriting word recognition. A clear prompt is designed to guide the model in extracting word from image, while strictly limiting the output format.


To sum up, the main contributions of this study are as follows:
\begin{itemize}
\item We propose a modality mapping architecture, $m^2VLMs$, which serves as a bridge between mmWave data and VLMs, facilitating the transfer of the generalization capability of VLMs to RF sensing applications. To foster further exploration in this emerging field, we collect and release a 3D mmWave in-air handwriting dataset\footnote{\url{www.kaggle.com/datasets/mmpencil/mmpencil-dataset}} containing 7,664 samples of 200 different words, 12 users, and 22 writing scenarios. To the best of our knowledge, this is the first attempt to apply VLMs for RF-based handwriting recognition.

\item We design a trajectory reconstruction algorithm for 3D spatial handwriting movements, which effectively eliminates the influence of writing position on handwriting trajectories by extracting multidimensional spectral features. 

\item We design a spatial-to-visual mapping algorithm that converts mmWave data into image data, effectively eliminating the interference of writing style related factors (e.g., the speed and direction) by leveraging the powerful generalization capability of VLMs.

\item We develop a prototype handwriting recognition system named \textit{mmPencil} based on COTS mmWave radar and leading VLMs. Extensive experiments in real-world scenarios demonstrate that \textit{mmPencil} achieves accurate and robust word-level handwriting recognition, unaffected by varying writing styles and environmental factors. In particular, the recognition accuracy for 200 words stands at 97.60\%, with the zero-shot recognition accuracy remaining as high as 92.50\% for unseen users during training.
\end{itemize}

The rest of this paper is organized as follows. We present the related work in \S \ref{related work}, followed by an introduction to the preliminaries in \S \ref{preliminaries}. The detailed design of the $m^2VLMs$ architecture is given in \S \ref{details}. The technical details of trajectory reconstruction and data mapping are then presented in \S \ref{reconsturction} and \S \ref{mapping}, respectively. Finally, we evaluate the performance of \textit{mmPencil} in \S \ref{evaluation}, followed by discussion and conclusion in \S \ref{discussion} and \S \ref{conclusion}.


\section{Related Work} \label{related work}
Handwriting recognition is an important branch of gesture recognition \cite{xiao2021onefi,liu2024unifi,gao2023wicgesture,gao2022towards,gao2021towards,niu2021understanding,zheng2019zero,yu2022sodar,wang2020gesture,zhang2023wirite}. By capturing the precise trajectory of the in-air writing movements, it achieves a transition from coarse-grained action detection to fine-grained information representation in gesture recognition. We summarize a series of the most representative works in Table \ref{table: related work}.

\begin{table}[htb]
  \caption{Comparison with Existing RF-Based In-Air Writing Recognition Systems.}
  \label{table: related work}
  \small
  \begin{tabular}{cccccc}
    \toprule
    System & Radio & \makecell{Recognition Scope} & \makecell{Writing-Position \\ Independent} & \makecell{Writing-Direction \\ Independent} & \makecell{Writing-Order \\ Independent}\\
    \midrule
    RF-finger \cite{wang2018multi} & RFID & 26 letters \& 4 shapes & 2D plane & \XSolidBrush & \XSolidBrush \\
    Arsalan et al. \cite{arsalan2019character} & mmWave & 10 letters \& 5 numbers & 2D plane & \Checkmark & \Checkmark \\
    FingerDraw \cite{wu2020fingerdraw} & Wi-Fi & 26 letters \& 10 numbers & 2D plane & \XSolidBrush & \XSolidBrush \\
    mmWrite \cite{regani2021mmwrite} & mmWave & 50 characters & 2D plane & \XSolidBrush & \XSolidBrush \\
    CentiTrack-3D \cite{han2021towards} & Wi-Fi & 26 letters \& 10 numbers & \textbf{3D space} & \XSolidBrush & \XSolidBrush \\
    HandGest \cite{zhang2022handgest} & Wi-Fi & 12 letters \& 10 numbers & 2D plane & \XSolidBrush & \XSolidBrush \\
    IndexPen \cite{wei2022indexpen} & mmWave & 26 letters \& 5 gestures & \textbf{3D space} & \XSolidBrush & \XSolidBrush \\
    GWrite \cite{regani2022gwrite} & Wi-Fi & 15 letters & 2D plane & \XSolidBrush & \XSolidBrush \\
    MMHTSR \cite{chen2023mmhtsr} & mmWave & 26 letters \& 4 gestures & 2D plane & \XSolidBrush & \XSolidBrush \\
    AirWrite \cite{lin2024airwrite} & mmWave & 26 letters \& 5 words & 2D plane & \XSolidBrush & \XSolidBrush \\
    \bottomrule
    mmPencil & mmWave & \textbf{200 words} & \textbf{3D space} & \Checkmark & \Checkmark \\
    \bottomrule
\end{tabular}
\end{table}

In general, existing studies can be categorized into two groups, the first being model-driven methods that aim to establish a mapping between signal variations and handwriting actions. For example, the FingerDraw system \cite{wu2020fingerdraw} employs a CSI-quotient model to eliminate noise using the ratio of CSI signals between two antennas. This enables precise tracking of finger movements at sub-wavelength scales and, through signal segmentation and trajectory reconstruction algorithms, achieves two-dimensional (2D) trajectory reconstruction of finger movements. However, the limited bandwidth of Wi-Fi signals constrains FingerDraw' s performance in fine-grained finger-tracking tasks. The mmWrite system \cite{regani2021mmwrite} proposes a passive handwriting trajectory tracking method using a single mmWave radar. By leveraging advanced signal processing techniques, it achieves millimeter-level precision in handwriting trajectory reconstruction. The HandGest system \cite{zhang2022handgest} utilizes Dynamic Phase Vector (DPV) and Motion Rotation Variable (MRV) as gesture features, applying a hierarchical perception system for phased feature extraction and classification. This robust system enables finger tracking and air handwriting recognition without requiring large-scale data collection. The GWrite system \cite{regani2022gwrite} leverages CSI temporal features and time-reversal techniques, combining feature extraction and geometric analysis to achieve through-wall gesture recognition. However, its recognition capability for complex curved handwriting trajectories remains limited. The AirWrite system \cite{lin2024airwrite} introduces a handwriting recognition algorithm based on vector matrix generation and string matching. Without pre-training, it relies solely on simple string similarity calculations to identify handwriting trajectories. Nevertheless, this approach struggles with unseen letter writing-order, limiting its application in practical scenarios. Although these systems have achieved commendable results, they confine handwriting actions to 2D planes, whereas handwriting often occurs in 3D space. The CentiTrack-3D system \cite{han2021towards} addresses this limitation by integrating Angle of Arrival (AoA), Time of Flight (ToF), and variations in reflection path length to construct a high-precision 3D spatial gesture tracking model. However, its performance remains sensitive to the accuracy of initial position estimation, posing challenges for broader applications.

The other group is data-driven methods, which rely on large volumes of labeled data to train deep neural networks, enabling them to learn and extract unique patterns from hand movements. For example, the RF-finger system \cite{wang2018multi} utilizes gesture reflection features captured in three frames (corresponding to the initial, middle, and final phases of a gesture) and trains a CNN model to achieve fine-grained gesture recognition with high precision. Arsalan et al. \cite{arsalan2019character} propose an in-air writing recognition system based on multiple mmWave radar devices, which combines trajectory time series and 2D images, leveraging LSTM-CTC and DCNN models for character classification with notable performance. Focusing on two-finger character input, the IndexPen system \cite{wei2022indexpen} combines a hybrid architecture of CNN and LSTM to fully exploit the range-doppler and range-azimuth-elevation features of the mmWave radar and successfully realizes the effective extraction of spatial and temporal information. Similarly, the MMHTSR system \cite{chen2023mmhtsr} introduces an air-handwriting trajectory detection and reconstruction system, utilizing advanced signal processing techniques and Bidirectional-LSTM to achieve high-accuracy handwriting trajectory recognition and gesture classification. However, these CNN- and LSTM-based methods face limitations in generalizing to unseen complex writing movements and adapting to user-varying writing style preferences, significantly hindering their applicability in real-world scenarios.

In contrast, \textit{mmPencil} combines the advantages of model-driven and data-driven methods. It employs a multi-stage spatial trajectory reconstruction algorithm that effectively eliminates the influence of factors such as writing position, direction, and speed. Additionally, by leveraging a spatial-to-visual mapping algorithm, it capitalizes on the ability of VLMs to perceive the physical world. Even in the presence of significant varying of handwriting styles and environmental factors, \textit{mmPencil} maintains a high sensing performance.

\section{Preliminaries} \label{preliminaries}
In this section, we will introduce the preliminaries that form the basis for designing the proposed architecture and algorithms. Specifically, we first delve into the principles of multiple-input multiple-output (MIMO) antenna arrays based on time-division multiplexing (TDM). Subsequently, we explore the key techniques for range estimation and multi-dimensional angle estimation.

\begin{figure}[tb]
	\centering
	\begin{minipage}[t]{0.45\textwidth}
		\centering
  		\includegraphics[width=\linewidth]{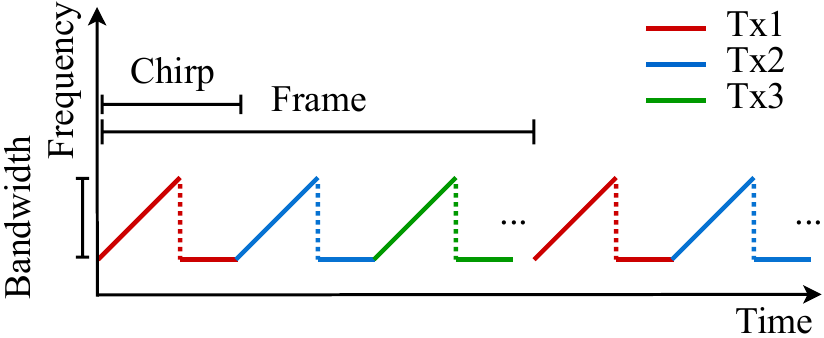}
  		\caption{mmWave radar Tx antennas alternately emit signals under TDM-MIMO.}
  		\label{fig: TDM}
	\end{minipage}\hspace{5mm}
	\begin{minipage}[t]{0.45\textwidth}
		\centering
  		\includegraphics[width=\linewidth]{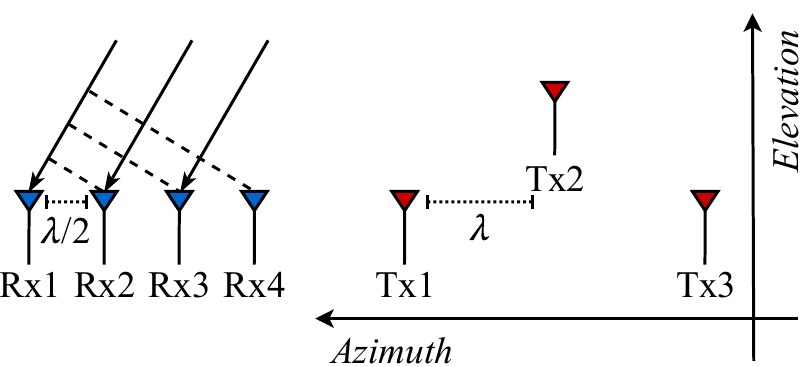}
  		\caption{MIMO antenna array of mmWave radar.}
  		\label{fig: MIMO}
	\end{minipage}
\end{figure}

\subsection{Antenna Array}

Unlike traditional pulse radars, mmWave radars emit signals with a frequency that varies linearly over time, thereby referred to as Frequency-Modulated Continuous Wave (FMCW) radars. By analyzing the ToF and Doppler shift of the received signals, FMCW radars can simultaneously measure the range and velocity of a target. However, relying solely on range and velocity information is insufficient to distinguish multiple targets. To address this limitation, FMCW radars employ MIMO antenna arrays, which enable spatial separation of targets through angle estimation. 

The core concept of MIMO antenna arrays lies in combining the signal paths of multiple transmit (Tx) and receive (Rx) antennas to form an equivalent virtual antenna array. The number of antennas in this virtual array is equal to the product of the physical Tx and Rx antennas. This design significantly enhances the angle resolution and target detection capability of radars without a substantial increase hardware costs. As shown in Fig. \ref{fig: TDM}, the transmit antennas emit signals in chronological order in TDM mode. The receive antennas capture the echo signals reflected from the targets. By jointly processing the received signals, target signals can be effectively separated across both temporal and spatial dimensions.

\subsection{Range Estimation}

The transmit signal of an FMCW radar exhibits linear frequency modulation over time, mathematically expressed as:
\begin{equation}
    S_{Tx}(t) = A_{Tx}\cos\left(2\pi\left(f_{0}t + \frac{B t^{2}}{2T}\right) + \varphi_{Tx}\right), 
\end{equation}
where $A_{Tx}$ is the amplitude of the transmit signal, $f_0$ is the starting frequency, $B$ is the signal bandwidth, $T$ is the sweep time, and $\varphi_{Tx}$ is the random initial phase. After the transmit signal is reflected by the target, the received signal is a time-delayed version of the transmit signal. By processing the transmit and received signals through a mixer, the intermediate frequency (IF) signal can be obtained as:
\begin{equation}
     S_{IF}(t) = A_{IF}\cos\left(2\pi\left(f_{0}\tau + \frac{B}{2T}\left(2t\tau - \tau^2\right)\right) + \Delta \varphi\right),    
\end{equation}
where $\tau$ is the signal propagation delay, and $\Delta \varphi = \varphi_{Tx} - \varphi_{Rx}$ is the phase difference, which is typically negligible. 

The relationship between the target range $R$ and the frequency $f$ of the IF signal is given by:
\begin{equation}
    R = \frac{CTf}{2B},
\end{equation}
where $C$ is the speed of light. By performing a Fast Fourier Transform (FFT) on the received signal, range information can be extracted. The range resolution of the radar is limited by the signal bandwidth and is given by:
\begin{equation}
    R_{res} = \frac{C}{2B}.
\end{equation} 
For example, an FMCW radar with a bandwidth of 4 GHz can achieve a range resolution of approximately 3.75 cm. However, in practical scenarios, factors such as hardware limitations and environmental noise must also be considered.

\subsection{Multi-dimensional Angle Estimation}

Angle estimation is achieved by analyzing the spatial characteristics of the target signals captured by the receiving antenna array. In traditional one-dimensional angle estimation, mmWave radars primarily measure the azimuth angle of targets. The azimuth AoA $\theta$ can be expressed as:
\begin{equation}
    \theta = \arcsin\left(\frac{\lambda \Delta \Phi}{2\pi d}\right),
\end{equation}
where $d$ is the receiving antenna spacing, $\lambda$ is the wavelength, $d\sin(\theta)$ and $\Delta \Phi$ are the path difference between neighboring receiving antennas and the resulting phase difference, respectively. By performing an angle-FFT on the received signals, the azimuth angles of multiple targets can be resolved. Furthermore, increasing the number of receiving antennas $n_{Rx}$ effectively improves the angle resolution, which can be expressed as:
\begin{equation}
    \theta_{res} = \arcsin\left(\frac{\lambda}{n_{Rx} d \cos(\theta)}\right).
\end{equation}

In modern mmWave radars, the detection of elevation angles can be achieved by adjusting the layout of the transmit antennas, particularly by introducing height differences. As shown in Fig. \ref{fig: MIMO}, the use of transmit antennas at different heights (i.g., Tx2) enables the estimation of elevation angles by leveraging the vertical path difference. Specifically, the path difference in the vertical direction introduces $h \sin(\phi)$, where $h$ represents the height difference and $\phi$ is the elevation angle, calculated as $\phi = \arcsin\left(\frac{\lambda \Delta \Phi_{el}}{2\pi h}\right)$, where $\Delta \Phi_{el}$ denotes the phase difference caused by the vertical path difference.

By combining azimuth and elevation angles using 2D angle estimation, mmWave radars can achieve 3D target detection. This capability makes it feasible to extend applications such as in-air writing recognition from a 2D plane to a 3D space.

\section{Overview of the $m^2VLMs$ Architecture} \label{details}
\begin{figure}
	\centering
	\begin{minipage}[htbp]{\textwidth}
		\centering
  		\includegraphics[width=0.85\linewidth]{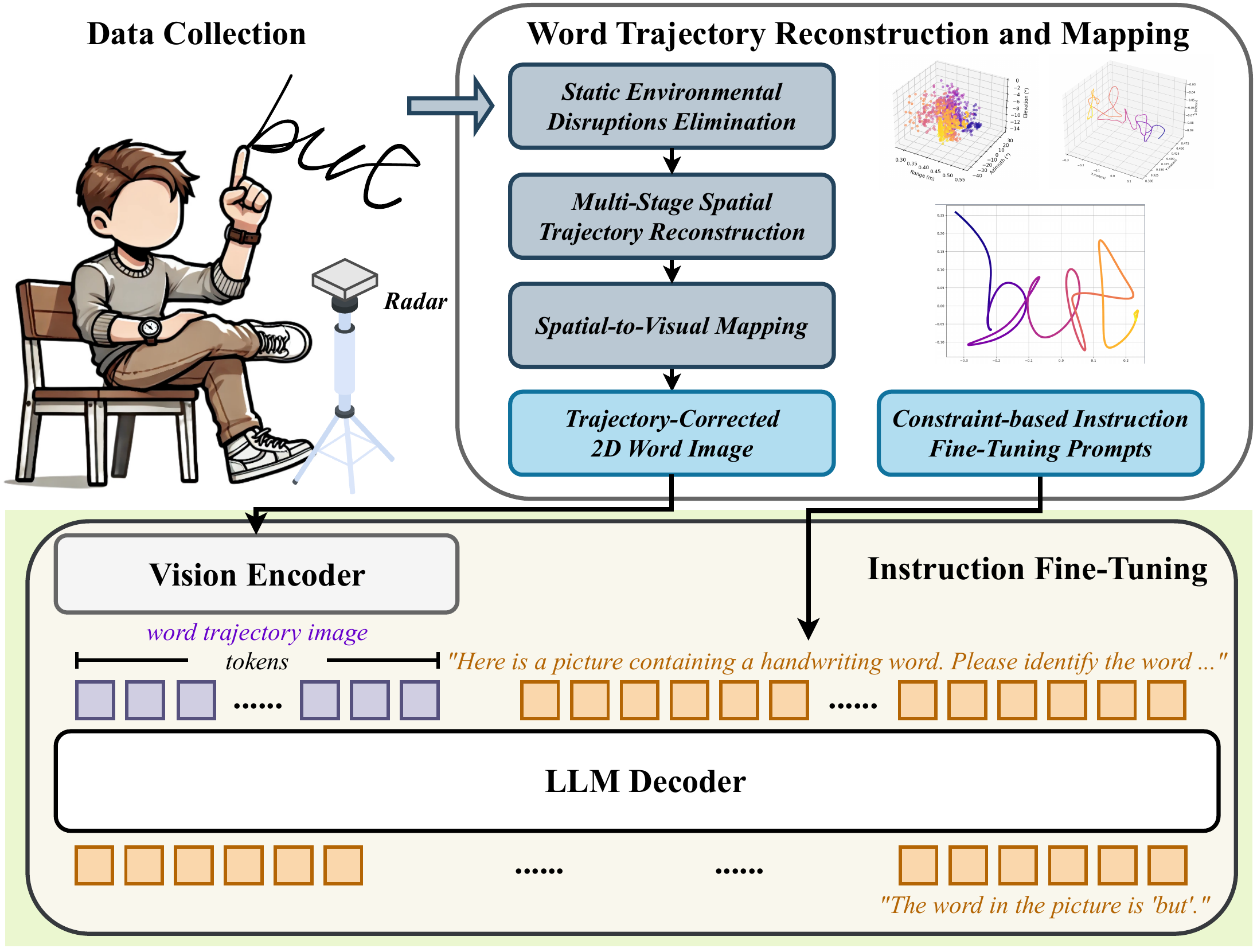}
  		\caption{Overview of the $m^2VLMs$ architecture.}
  		\label{fig: system}
	\end{minipage}
\end{figure}

Due to the abundance of large-scale image-text data, VLMs possess outstanding generalization capabilities in sensing and understanding the physical world. To unlock this great potential of VLMs in the domain of RF sensing, we propose a modality mapping architecture that connects mmWave data and VLMs, as shown in Fig. \ref{fig: system}. Specifically, the proposed $m^2VLMs$ architecture consists of four modules: data collection, trajectory reconstruction, data mapping, and instruction fine-tuning. In particular, the trajectory reconstruction module effectively eliminates the influence of environmental factors and writing position, while the data mapping module mitigates the impact of writing order and direction. We will provide detailed descriptions of these two modules in \S \ref{reconsturction} and \S \ref{mapping}, respectively.

\textbf{\textit{Data Collection.}} $m^2VLMs$ is built on a commercial off-the-shelf (COTS) 60 GHz mmWave radar. The radar is equipped with a MIMO antenna array comprising 3 transmitting antennas and 4 receiving antennas, offering a $120^\circ$ azimuth field of view and a $60^\circ$ elevation field of view. We place the radar horizontally on a desktop or similar platform to capture the writing movements.

\textbf{\textit{Trajectory Reconstruction.}} We design a set of trajectory reconstruction methods that effectively reduces the impact of variations in writing position on air handwriting recognition. First, we calculate the vector mean of time-domain data to suppress environmental clutter caused by multi-path effects. Next, we propose a multi-stage spatial trajectory reconstruction algorithm. Using range-azimuth and range-elevation profiles after spectrum refinement, we dynamically extract the center coordinates of target clusters, accurately capturing the spatial (i.e., range, azimuth, and elevation) trajectory at each moment. The obtained trajectory sequences are refined through outlier removal, followed by smoothing with a 3D Kalman filter.

\textbf{\textit{Data Mapping.}} We design a spatial-to-visual mapping algorithm. Through direction-corrected PCA, we map 3D trajectories to 2D images while preserving the shape characteristics of handwriting words. Image-based methods have high sequential robustness, which can mitigate the differences caused by varying writing directions and styles (i.e., order) across different users in time series. This mapping algorithm bridges the gap between RF signals and the visual images, enabling VLMs to interpret and process wireless signals with greater intuitiveness.

\textbf{\textit{Instruction Fine-Tuning.}} We design an instructional prompt to clearly guide the VLM in extracting text from images while strictly constraining the output format. The content of the prompt is as follows: \textit{"<image> Here is a picture containing a handwriting word. Please identify the word in the image. Ensure that your output is a correctly spelled English word. Note: Only provide the word itself, without any additional content or explanation."} Using clear instructions, we define the VLM's task objective as accurately recognizing handwriting words from images. Additionally, we incorporate the principles of constraint-based prompts \cite{zhang2024large}, strictly limiting the output to \textit{"a single correctly spelled English word"} and excluding unnecessary content or explanations. This design combines task instructions with output constraints, effectively helping the VLM adapt to handwriting-based text recognition tasks and improving the simplicity and accuracy of the output.

\section{Writing Trajectory Reconstruction} \label{reconsturction}

\subsection{Environmental Disruption Elimination}

Due to the multi-path effect and Tx-Rx coupling \cite{stove1992linear}, the raw IF signals of a mmWave radar are impacted by multiple factors, often including irrelevant components like direct leakage and environmental reflections. Therefore, to accurately extract the handwriting-related components from the composite signals, it is essential to preprocess the raw IF signals. While reference signal-based methods can effectively eliminate environmental noise, they often require additional collection of no-target data \cite{wang2024liqdetector}. 

\begin{figure}
	\centering
	\begin{minipage}[t]{\textwidth}
		\centering
  		\subfigure[Effectiveness of noise elimination.]{
  		\includegraphics[width=0.45\linewidth]{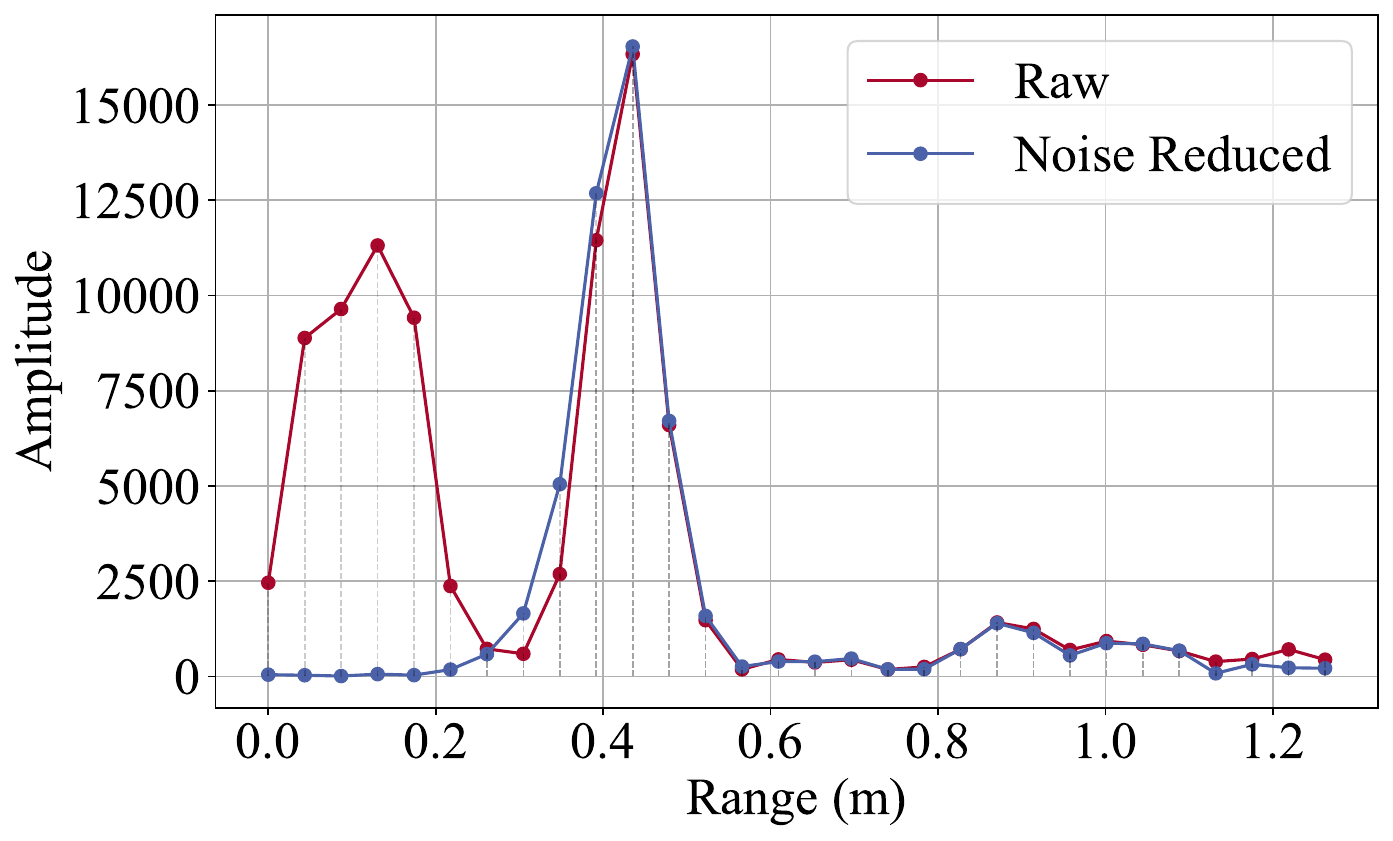}
  		\label{fig: Noise Elimination}}\hspace{10mm}
  		\subfigure[Effectiveness of spectrum refinement.]{
  		\includegraphics[width=0.45\linewidth]{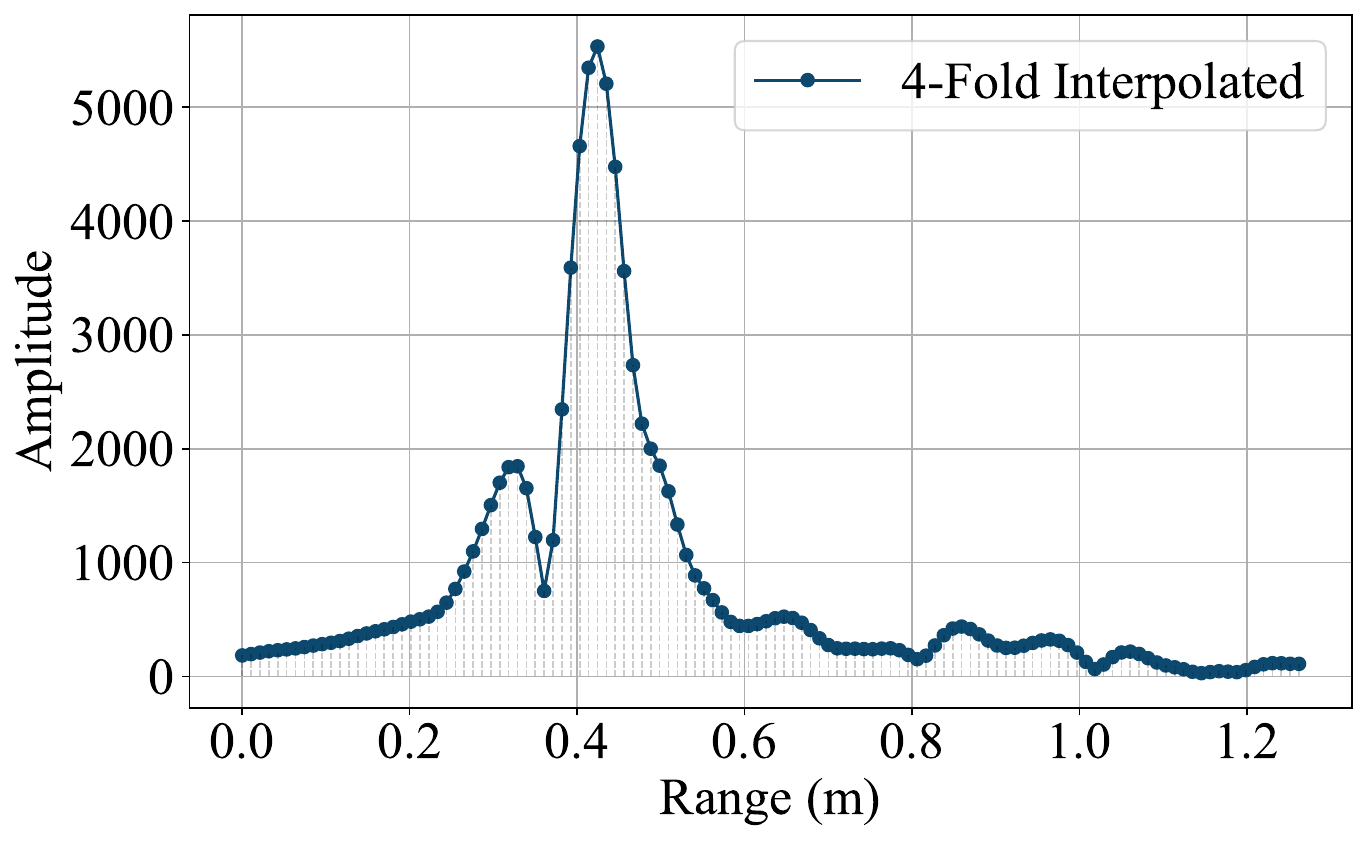}
  		\label{fig: Interpolating FFT}}
  	\end{minipage}
\caption{Illustration of data pre-processing results. (a) The noise elimination unit filters the interference of direct leakage and ambient noise. (b) The spectrum refinement unit improves the position estimation resolution.}
\end{figure}

To address this issue, we propose taking the mean of the time-domain signals as a reference and subtracting it from the raw IF signals at all times. Accordingly, the channel impulse response (CIR) at a specific time $t$ can be expressed as:
\begin{equation}
    h(\tau)=h(\tau,t)-\frac{1}{m}\sum^{m}_{i=1}h(\tau,t_i),
\end{equation}
where $m$ is the total number of frames in a complete handwriting signal. To provide a clearer demonstration, we collect data with the hand placed approximately 40 cm in front of the radar. Then, range-FFT is conducted on both the raw data and the noise-reduced data, and the results are illustrated in Fig. \ref{fig: Noise Elimination}. Since direct leakage (within 20 cm) and environmental noise (beyond 100 cm) tend to remain constant throughout the handwriting process, subtracting the mean vector can effectively eliminate these non-motion components.

\subsection{Multi-Stage Spatial Trajectory Reconstruction}
As we know, a FMCW radar is capable of detecting the relative range, azimuth and elevation angles of the target. Furthermore, based on the multi-stage spatial trajectory reconstruction algorithm, we can not only effectively suppress environmental interference, but also accurately estimate the finger position at each moment with super resolution. As a result, handwriting word recognition can be achieved regardless of the writing position.

\begin{figure}
	\centering
	\begin{minipage}[ht]{\textwidth}
		\centering
  		\subfigure[Range-azimuth profile.]{
  		\includegraphics[width=0.45\linewidth]{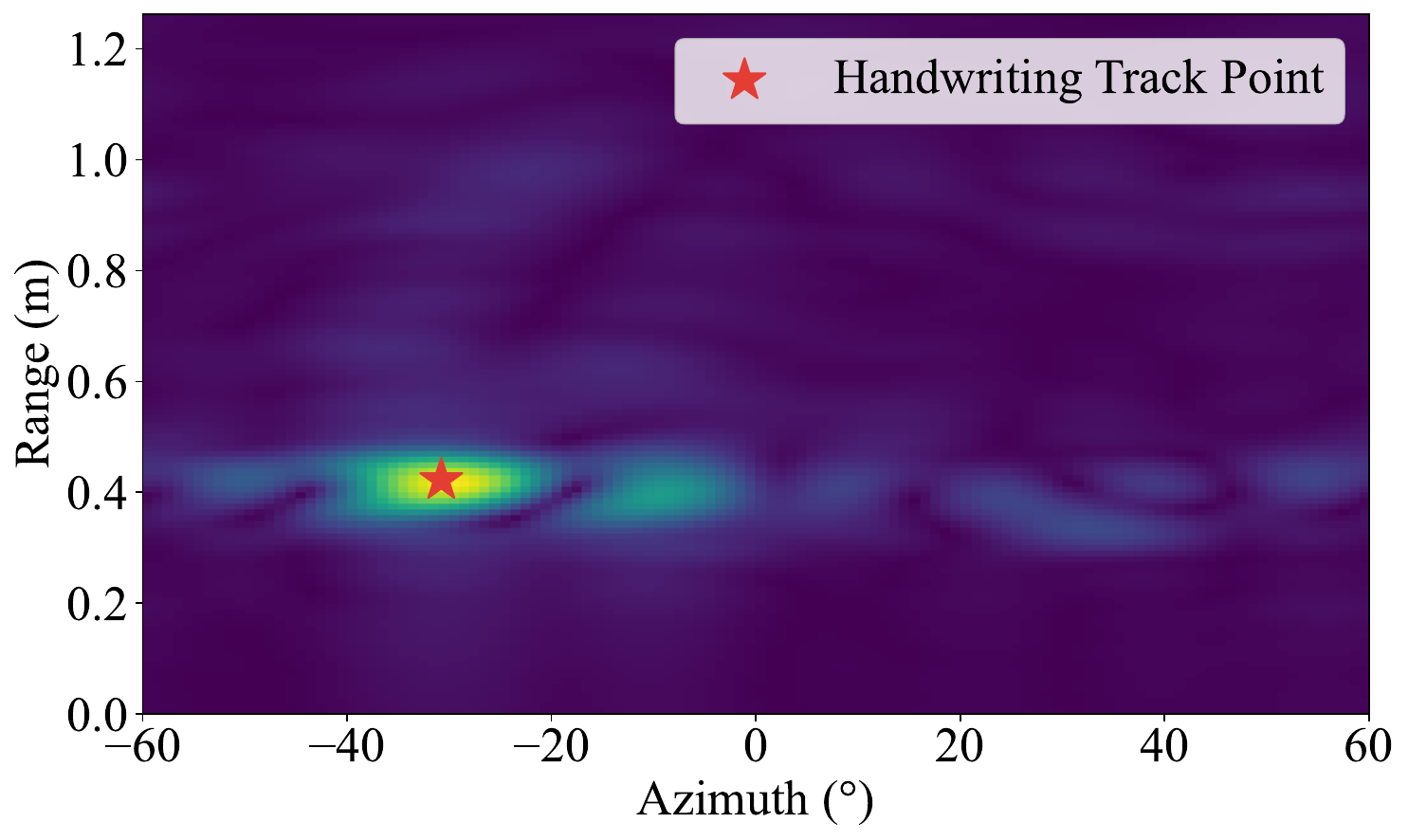}
  		\label{fig: RAM}}\hspace{10mm}
  		\subfigure[Range-elevation profile.]{
  		\includegraphics[width=0.45\linewidth]{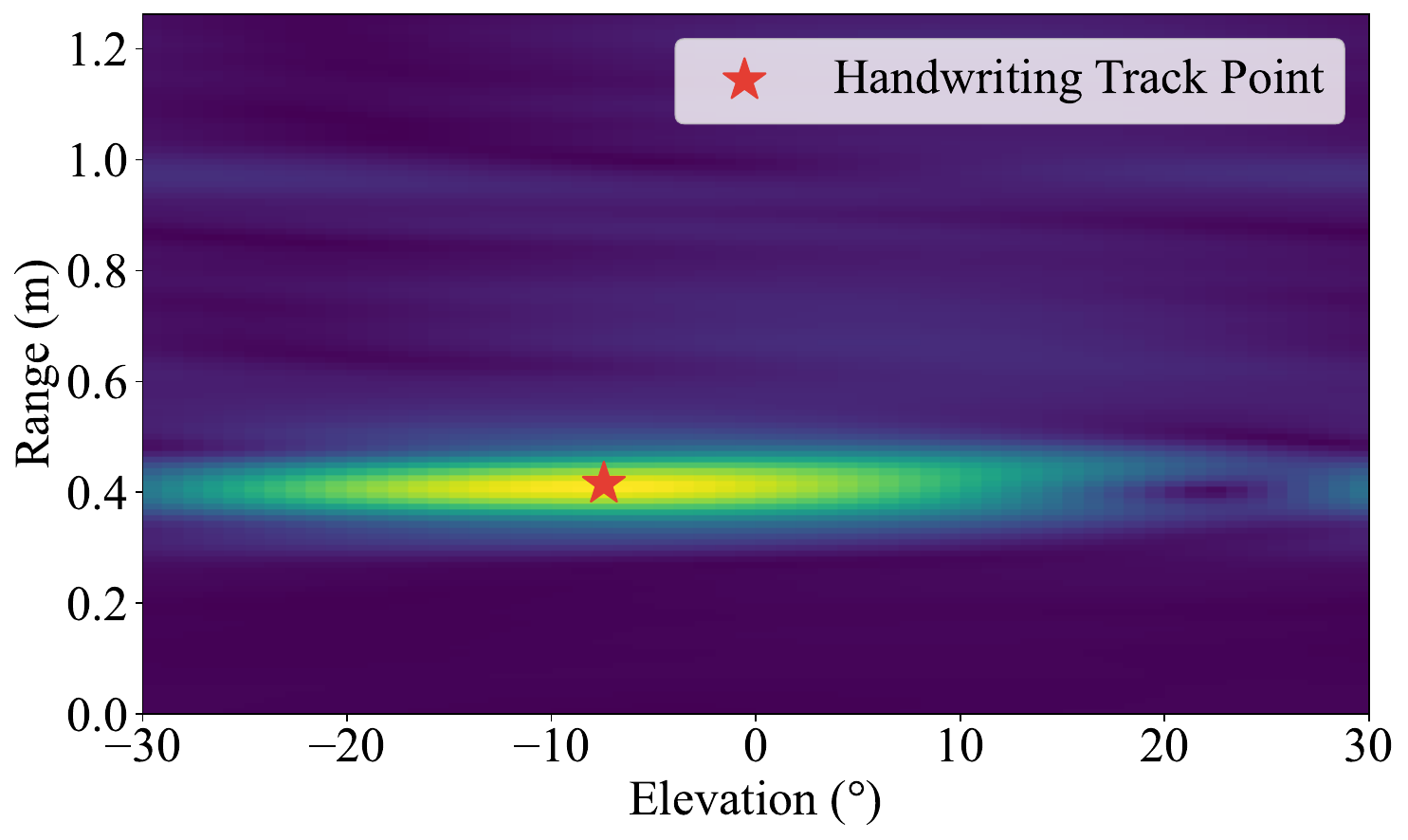}
  		\label{fig: REM}}
  	\end{minipage}
\caption{Illustration of spectrum refinement results.}
\end{figure}

\subsubsection{Range-Angle Spectral Refinement}
Generally, performing FFT on the fast-time dimension yields the range spectrum, where peaks represent reflective targets at specific ranges within the field of view. Further FFT processing on the channel dimension produces the range-angle profile. Notably, selecting data from Tx-Rx channels at the same elevation but different horizontal positions in the virtual antenna array (e.g., signals transmitted by Tx1 and Tx3 and received by the four Rx channels in Fig. \ref{fig: MIMO}) provides azimuth information. This is because the difference in reflected signals between these channels mainly arises from horizontal path differences. Similarly, elevation information can be extracted from Tx-Rx channels at the same horizontal position but different elevations (e.g., channels Tx2-Rx1 and Tx1-Rx3). Given the noise-canceled IF signal as $S(t)$ and the range as $r$, the range-azimuth and range-elevation profiles can be expressed as:
\begin{equation}
    RA(r, \theta, t) = \left| \text{FFT}_\theta(\text{FFT}_r(S(t))) \right|, \quad
    RE(r, \phi, t) = \left| \text{FFT}_\phi(\text{FFT}_r(S(t))) \right|.
\end{equation}
However, the resolution of FFT is limited by the sampling frequency, as constrained by the Nyquist Sampling Theorem \cite{vaidyanathan2001generalizations}. Larger sampling intervals introduce a picket fence effect, increasing the likelihood of missing true frequencies between CIR samples. For example, a commonly used 12-channel mmWave radar with a 4 GHz bandwidth can only achieve a range resolution of approximately 3.75 cm, an azimuth angle resolution of $15^\circ$, and an elevation angle resolution of $30^\circ$, which is far from sufficient for precise handwriting recognition tasks.

To address the motion signal overlap caused by insufficient radar resolution, we propose up-sampling the existing CIR samples by adding zero padding \cite{dhekne2018liquid,wu2020msense}. This method effectively inserts more frequency points in the spectrum without increasing the hardware sampling rate, improving the overall spectral resolution. Specifically, given the original signal $S_{\text{raw}}(t)$ of length $N_{\text{sample}}$, zero padding extends it to $N_{\text{interp}}$ points. The zero-padded signal can be expressed as:
\begin{equation}
    S_{\text{interp}}(t) =
    \begin{cases}S_{\text{raw}}(t), & 0 \leq t < N_{\text{sample}}, \\
    0, & N_{\text{sample}} \leq t < N_{\text{interp}}.
    \end{cases}
\end{equation}
The range-FFT with zero padding is then expressed as:
\begin{equation}
    S_{\text{FFT,range}}(k) = \sum_{t=0}^{N_{\text{interp}}-1} S_{\text{interp}}(t) \cdot e^{-\frac{j2\pi kt}{N_{\text{interp}}}}, \quad 0 \leq k < N_{\text{interp}}.
\end{equation}
where $k$ represents the $k$-th frequency point. Although zero padding does not change the physical range resolution, it increases the spectral density, allowing for more precise target localization. The refined resolution can be expressed as $R = \frac{C F_s}{2 N_{\text{interp}} S}$, where $F_s$ is the sampling rate and $S$ is the frequency modulation slope. 

Given $F_s = 2000 \, \text{ksps}$, $S = 66 \, \text{MHz}/\mu\text{s}$, and $N_{\text{interp}} = 108 \times 4$ (4x interpolation), the range resolution is approximately 1.05 cm, as shown in Fig. \ref{fig: Interpolating FFT}. the range estimate from the 4x interpolated FFT is 42.43 cm, while the original FFT estimates it as 43.53 cm. This discrepancy arises because the original FFT restricts the detection results to discrete sampling points, such as 39.18 cm and 43.53 cm. Similarly, spectral refinement of angle information can achieve an azimuth and elevation resolution of $1^\circ$, as shown in Fig. \ref{fig: RAM} and \ref{fig: REM}.

\subsubsection{Frequency-Domain Feature Extraction}
Handwriting movements are complex movements performed collaboratively by the arm and hand, resulting in target reflections distributed across multiple range and angle bins with significant spatial spread. Consequently, directly identifying the peak values from range-azimuth and range-elevation profiles as handwriting trajectory points may neglect the distributional characteristics of the motion, leading to inaccurate trajectory localization. To address this issue, we propose a method based on target point extraction and cluster center computation for more reliable handwriting trajectory determination.

\begin{figure}[t]
    \centering
    \begin{minipage}[t]{\textwidth}
        \centering
        \subfigure[Trajectory points of "hello".]{
            \includegraphics[width=0.25\linewidth]{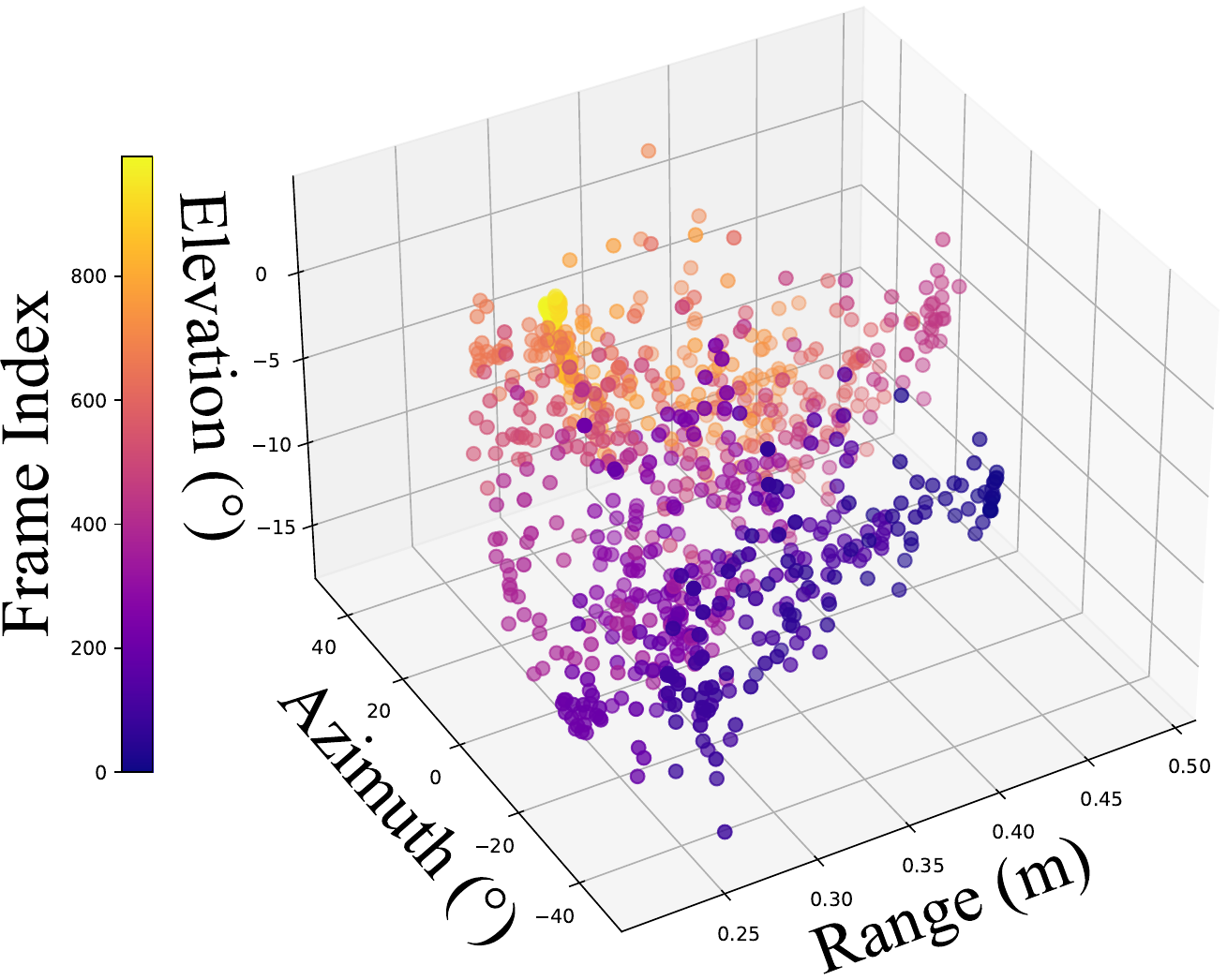}
            \label{fig: hello points}}
        \subfigure[Trajectory points of "world".]{
            \includegraphics[width=0.25\linewidth]{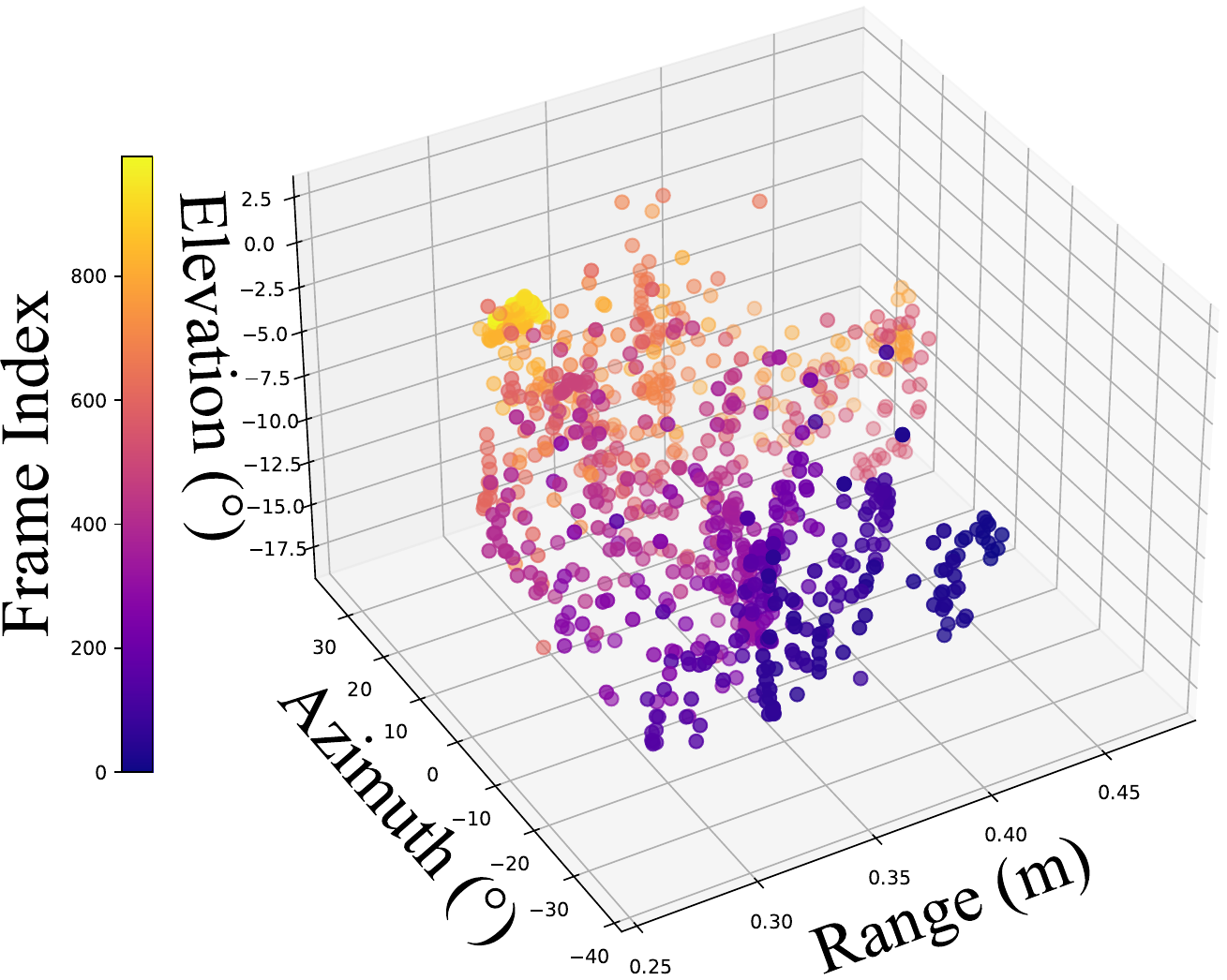}
            \label{fig: world points}}
        \subfigure[Smoothed trajectory of "hello".]{
            \includegraphics[width=0.22\linewidth]{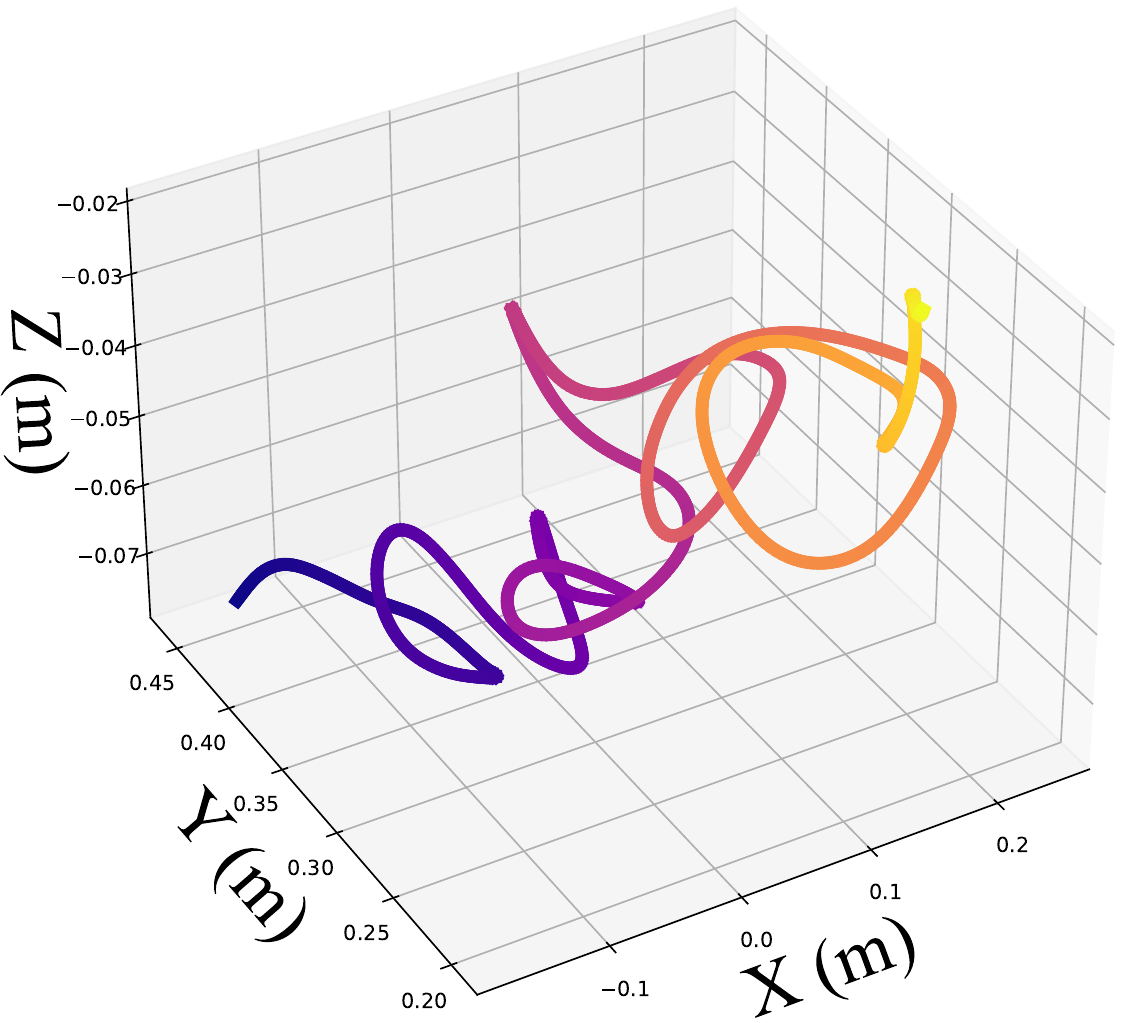}
            \label{fig: hello trajectory}}
        \subfigure[Smoothed trajectory of "world".]{
            \includegraphics[width=0.22\linewidth]{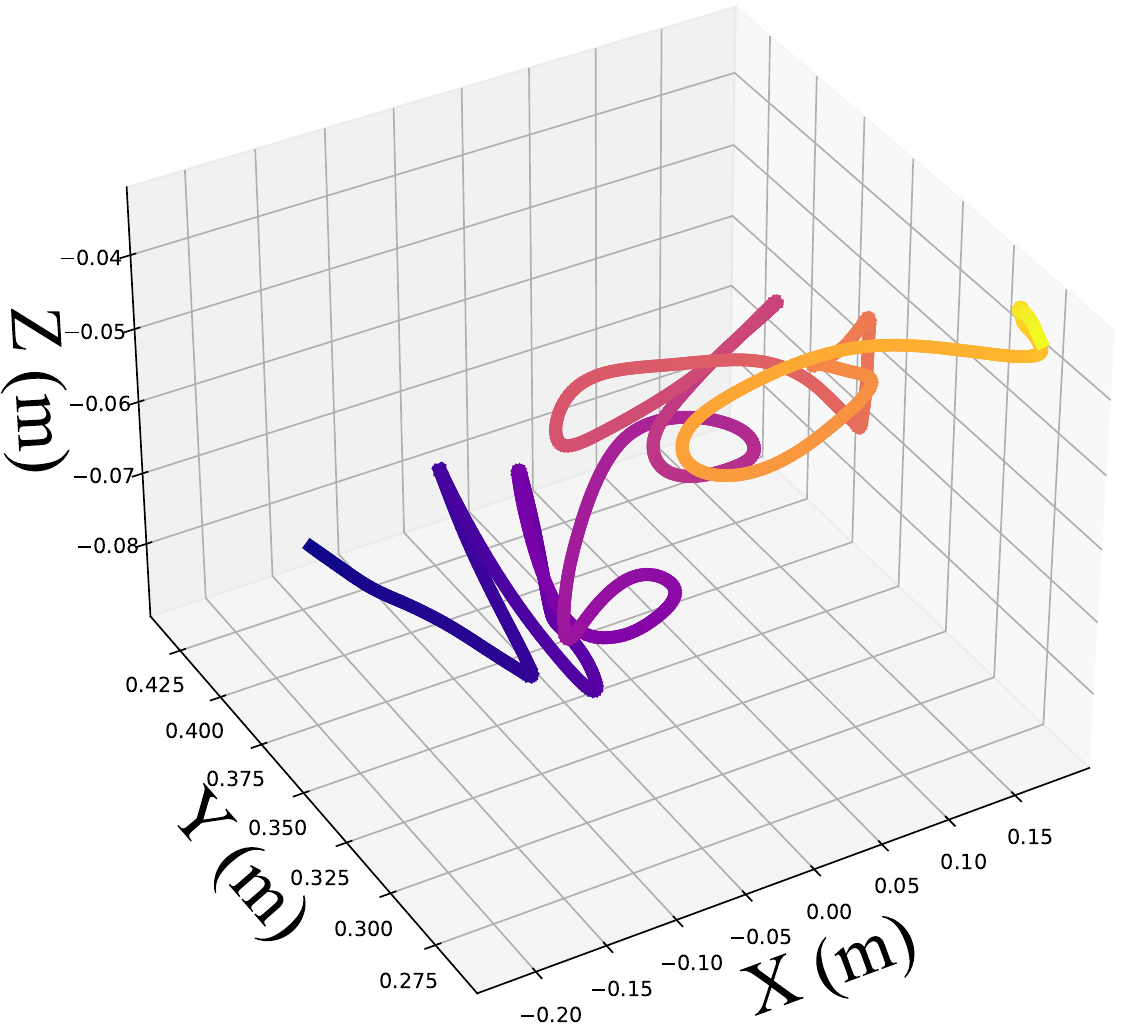}
            \label{fig: world trajectory}}
        \caption{Handwriting trajectory points in polar coordinates and smoothed handwriting trajectories in 3D Cartesian coordinates (The color change represents the writing order).}
    \end{minipage}
\end{figure}

Specifically, in each frame, we extract all target points exceeding a predefined threshold from the range-azimuth and range-elevation profiles. Let the signal strength at time $t_0$ be denoted as $M(r, \theta, t_0)$ and $M(r, \phi, t_0)$. The set of target points can be expressed as:
\begin{equation}
    P_{RA} = \{(r, \theta) \mid M(r, \theta, t_0) \geq \alpha \cdot \max(M) \}, \quad P_{RE} = \{(r, \phi) \mid M(r, \phi, t_0) \geq \alpha \cdot \max(M) \},
\end{equation}
where $\alpha$ is a scaling factor, we empirically set it to $0.75$ based on experience obtained during the data collection process. The cluster center of target points in a certain frame, as shown in Fig. \ref{fig: RAM} and \ref{fig: REM}, can be expressed as:
\begin{equation}
    (\bar{r}, \bar{\theta}) = \frac{1}{|P_{RA}|} \sum_{(r, \theta) \in P_{RA}} (r, \theta), \quad
    (\bar{r}, \bar{\phi}) = \frac{1}{|P_{RE}|} \sum_{(r, \phi) \in P_{RE}} (r, \phi).
\end{equation}
By calculating the cluster centers of the target point sets extracted from all time frames, we can estimate the positional information of the entire handwriting trajectory. As an example, we write the words "hello", and "world" in space. The complete trajectory points for each writing motion are shown in Fig. \ref{fig: hello points} and \ref{fig: world points}.

\subsubsection{Motion Segment Extraction and Trajectory Smoothing}

Integrating the overall motion of the arm and hand into trajectory point extraction fully leverages the distributional properties of hand reflection signals. However, unlike traditional handwriting on paper, in-air handwriting lacks clear starting and ending points. Preparatory actions (e.g., lifting the hand) and concluding actions (e.g., lowering the hand) during the handwriting process are also recorded by the mmWave radar as target point data. The motion trajectories of these actions often significantly differ from the actual handwriting trajectories, and directly including these data could disrupt the accuracy and continuity of the extracted handwriting trajectories. 

To address this issue, we perform outlier removal on the extracted handwriting trajectory time series to eliminate interference points unrelated to the handwriting process. Only trajectory points $|\mathbf{p}_i| = \{r_i, \theta_i, \phi_i\}$ that satisfy the following conditions are retained $|r_i - \mu_r| \leq k \cdot \sigma_r, |\theta_i - \mu_\theta| \leq k \cdot \sigma_\theta, |\phi_i - \mu_\phi| \leq k \cdot \sigma_\phi$, where $k$ is a predefined threshold multiplier, $\mu_r$, $\mu_\theta$, and $\mu_\phi$ represent the mean values of the 3D position information, and $\sigma_r$, $\sigma_\theta$, and $\sigma_\phi$ represent their respective standard deviations. After removing outliers, the resulting trajectory point time series better reflects the actual handwriting process while eliminating interference from non-writing movements.

Although spectrum refinement and trajectory point localization techniques are applied to enhance signal quality during data processing, the generated trajectory sequences inevitably exhibit fluctuations and high-frequency noise. These fluctuations result not only from the complexity of in-air writing but also from environmental disturbances and inherent uncertainties in signal processing.

\begin{figure}
	\centering
	\begin{minipage}[t]{\textwidth}
		\centering
  		\subfigure[transition covariance to $1 \times 10^{-2}$]{
  		\includegraphics[width=0.3\linewidth]{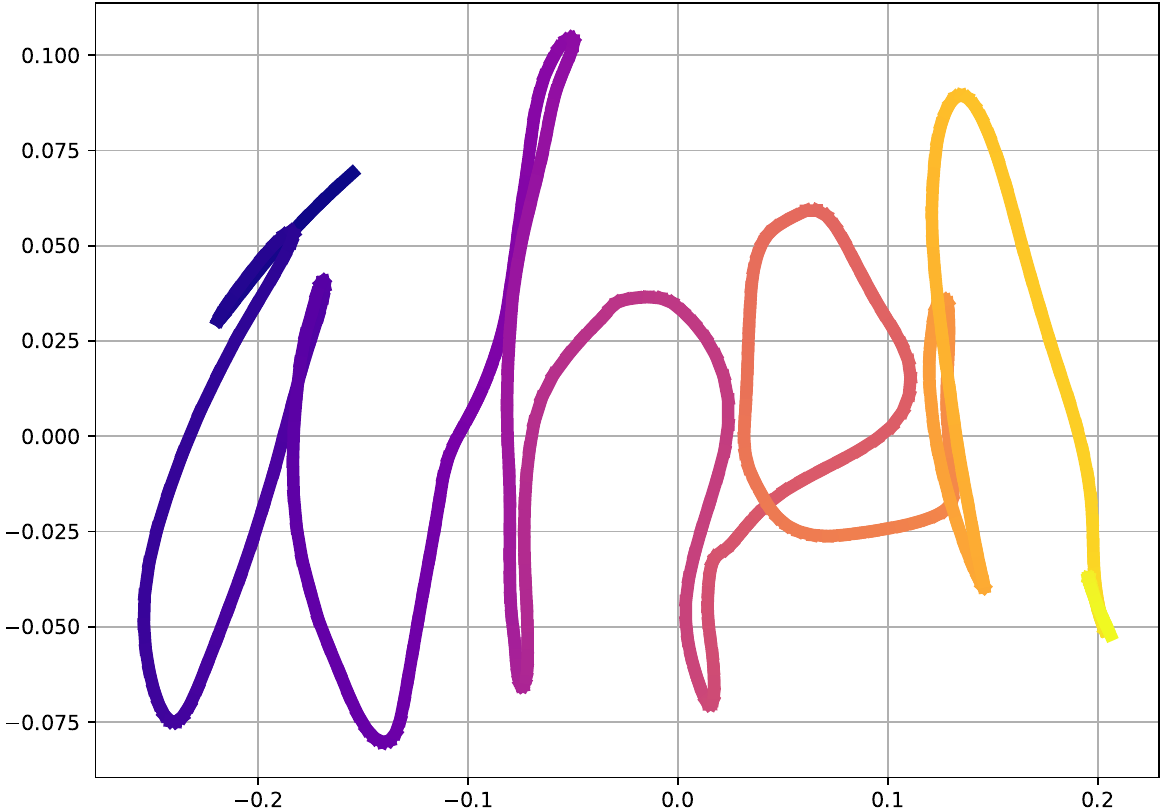}
  		\label{fig: 1e-2}}\hspace{3mm}
  		\subfigure[transition covariance to $1 \times 10^{-3}$]{
  		\includegraphics[width=0.3\linewidth]{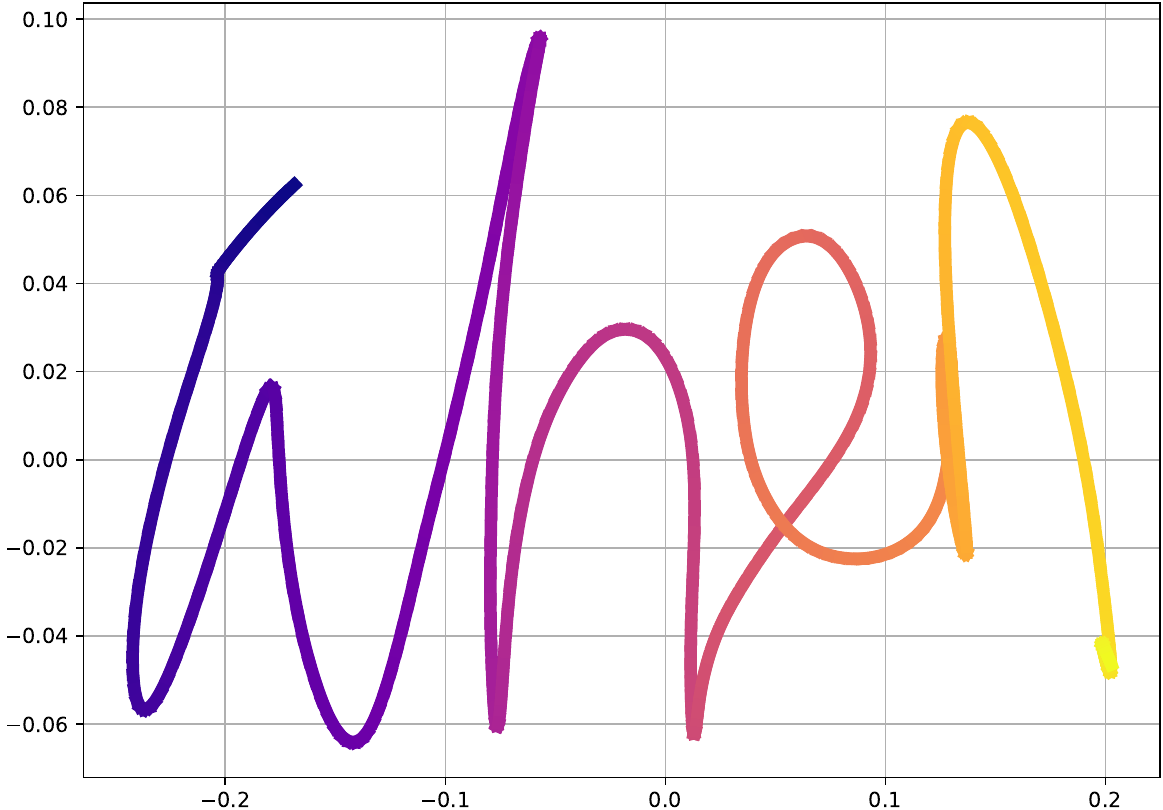}
  		\label{fig: 1e-3}}\hspace{3mm}
            \subfigure[transition covariance to $1 \times 10^{-4}$]{
  		\includegraphics[width=0.3\linewidth]{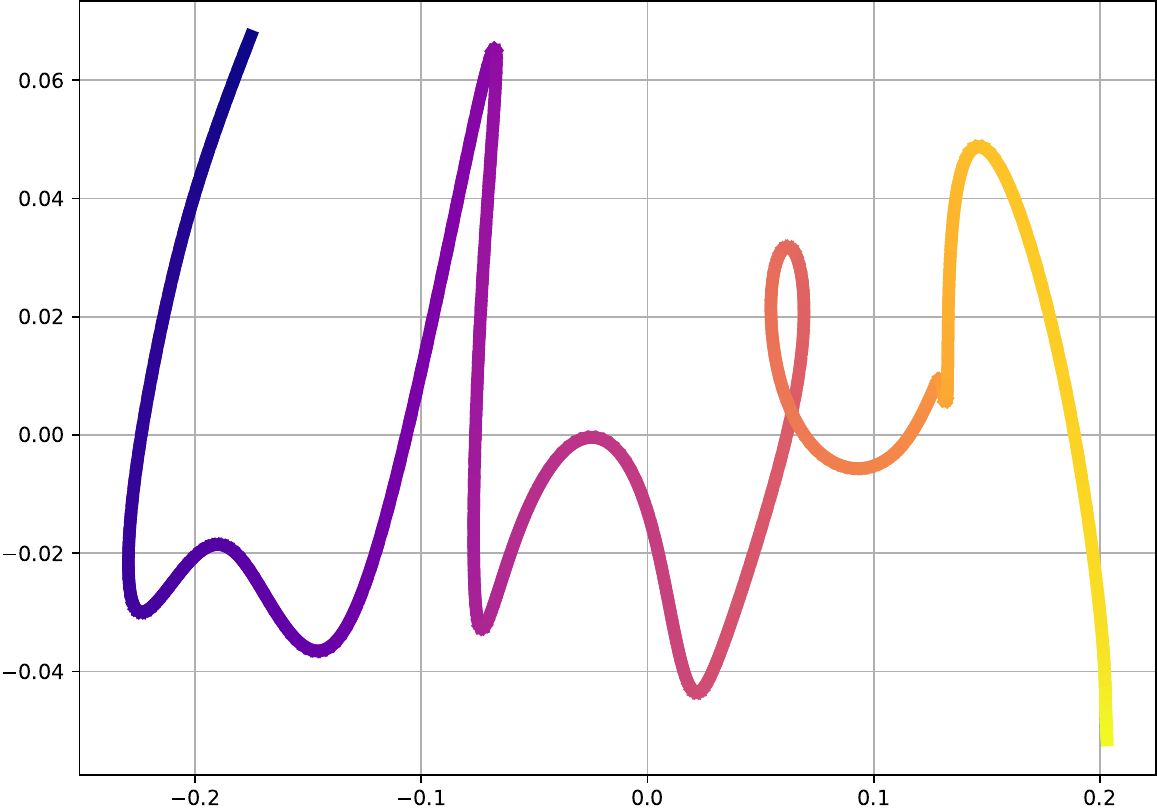}
  		\label{fig: 1e-4}}
  	\end{minipage}
        \vspace{-2mm}
\caption{Filtering effects with different transition covariance.}
\vspace{-3mm}
\label{fig: kalman}
\end{figure}

To optimize trajectory sequences, we introduce a 3D Kalman filter. Traditional filtering techniques, such as the Savitzky-Golay filter \cite{schafer2011savitzky}, rely on fixed window sizes to smooth data. However, these methods often cause detail loss when processing signals of varying lengths. This issue becomes particularly evident in character shape restoration. Additionally, windowing strategies distort periodic strokes, especially in letters like "o", "e", and "a", leading to compression or deformation during smoothing. To address these challenges, we adopt the dynamic prediction and update capabilities of the Kalman filter \cite{welch1995introduction}, and the smoothed trajectories of the example words are shown in Fig. \ref{fig: hello trajectory} and \ref{fig: world trajectory}. In our implementation of Kalman filtering, we typically set the observation covariance to $1$ and the transition covariance to $1 \times 10^{-3}$. As illustrated in Fig. \ref{fig: kalman}, the transition covariance significantly affects the quality of trajectory reconstruction. In particular, a smaller transition covariance tends to over-smooth the trajectory, thereby removing fine-grained details of the word writing process. In contrast, a larger transition covariance retains more details but also introduces considerable noise and clutter. This 3D Kalman filtering approach fully leverages the recursive computation of the Kalman filter, effectively smoothing the three-dimensional trajectory data while preserving the original shape of the characters and avoiding distortion issues common in traditional methods. 

\begin{figure}
	\centering
	\begin{minipage}[t]{\textwidth}
		\centering
  		\subfigure["but".]{
  		\includegraphics[width=0.23\linewidth]{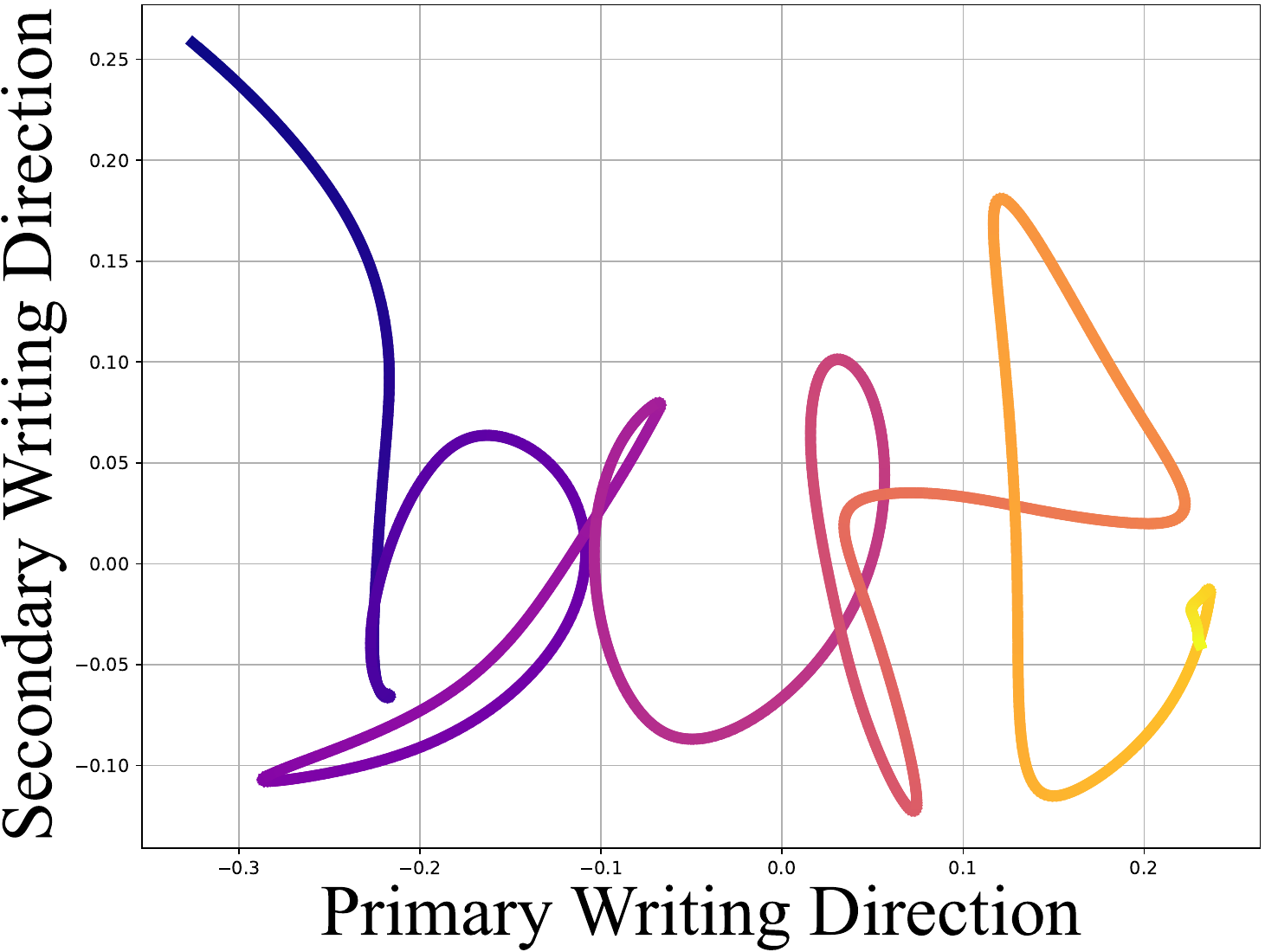}
  		\label{fig: but PCA}}
  		\subfigure["good".]{
  		\includegraphics[width=0.23\linewidth]{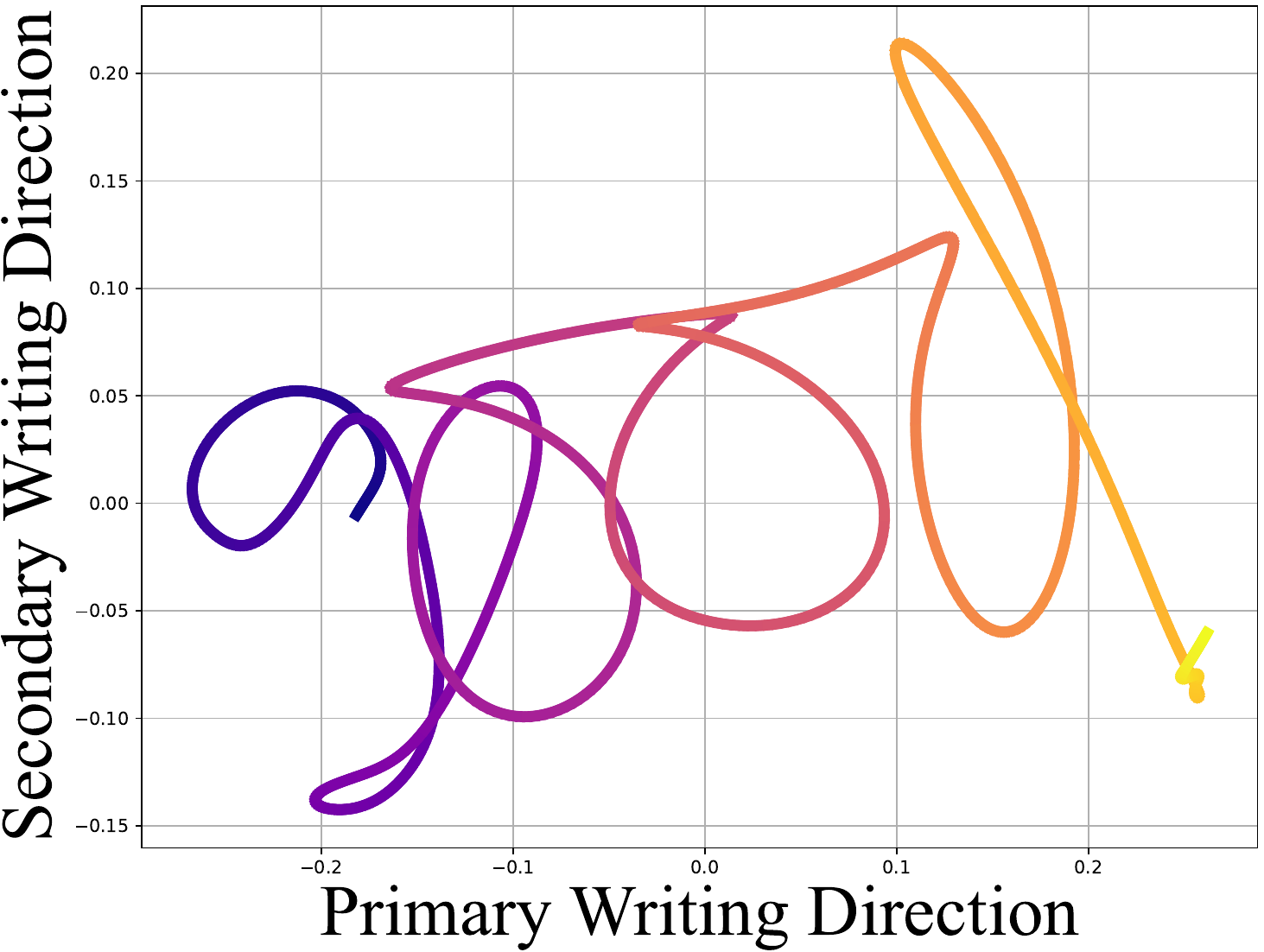}
  		\label{fig: good PCA}}
            \subfigure["hello".]{
  		\includegraphics[width=0.23\linewidth]{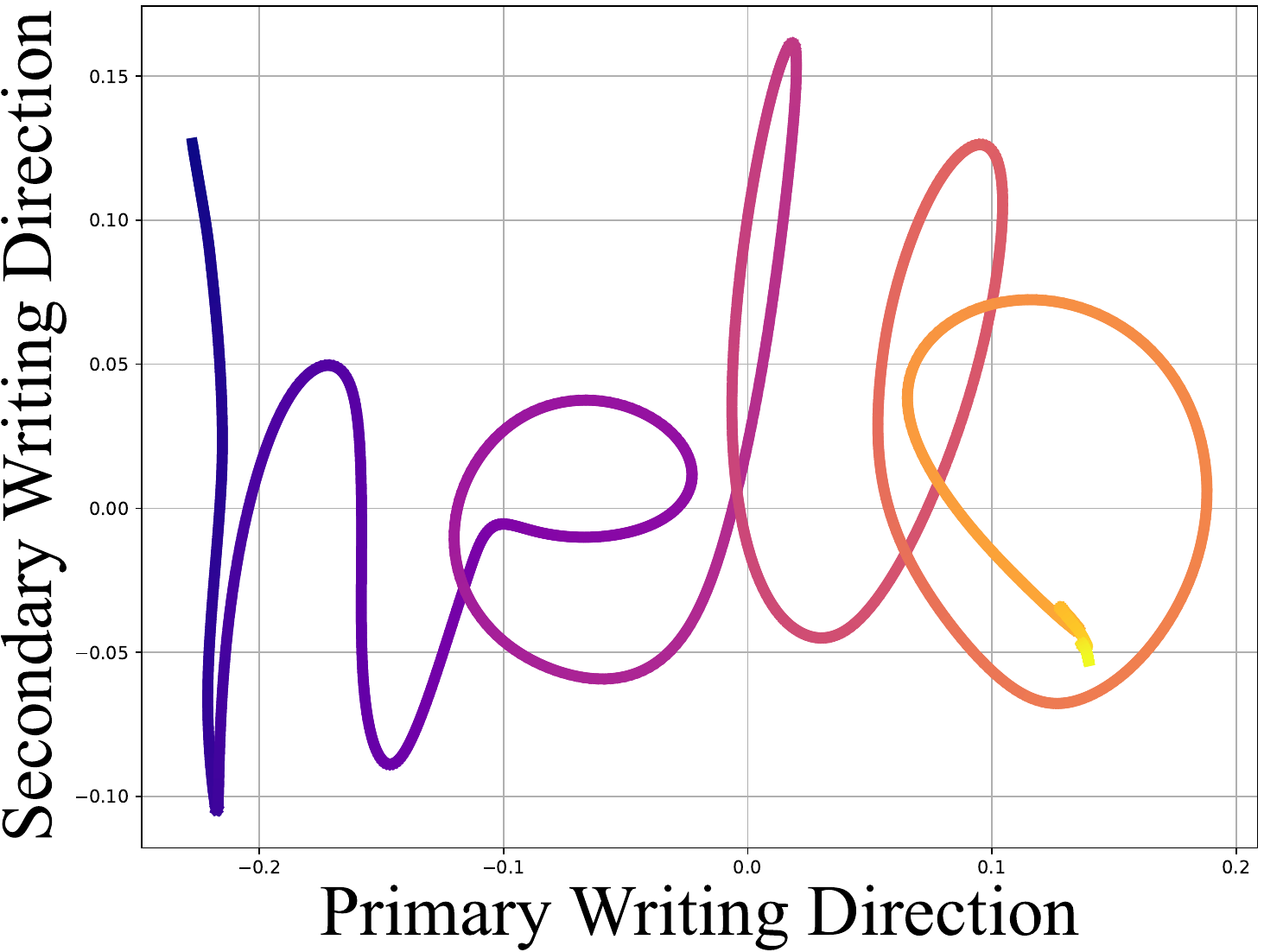}
  		\label{fig: hello PCA}}
            \subfigure["world".]{
  		\includegraphics[width=0.23\linewidth]{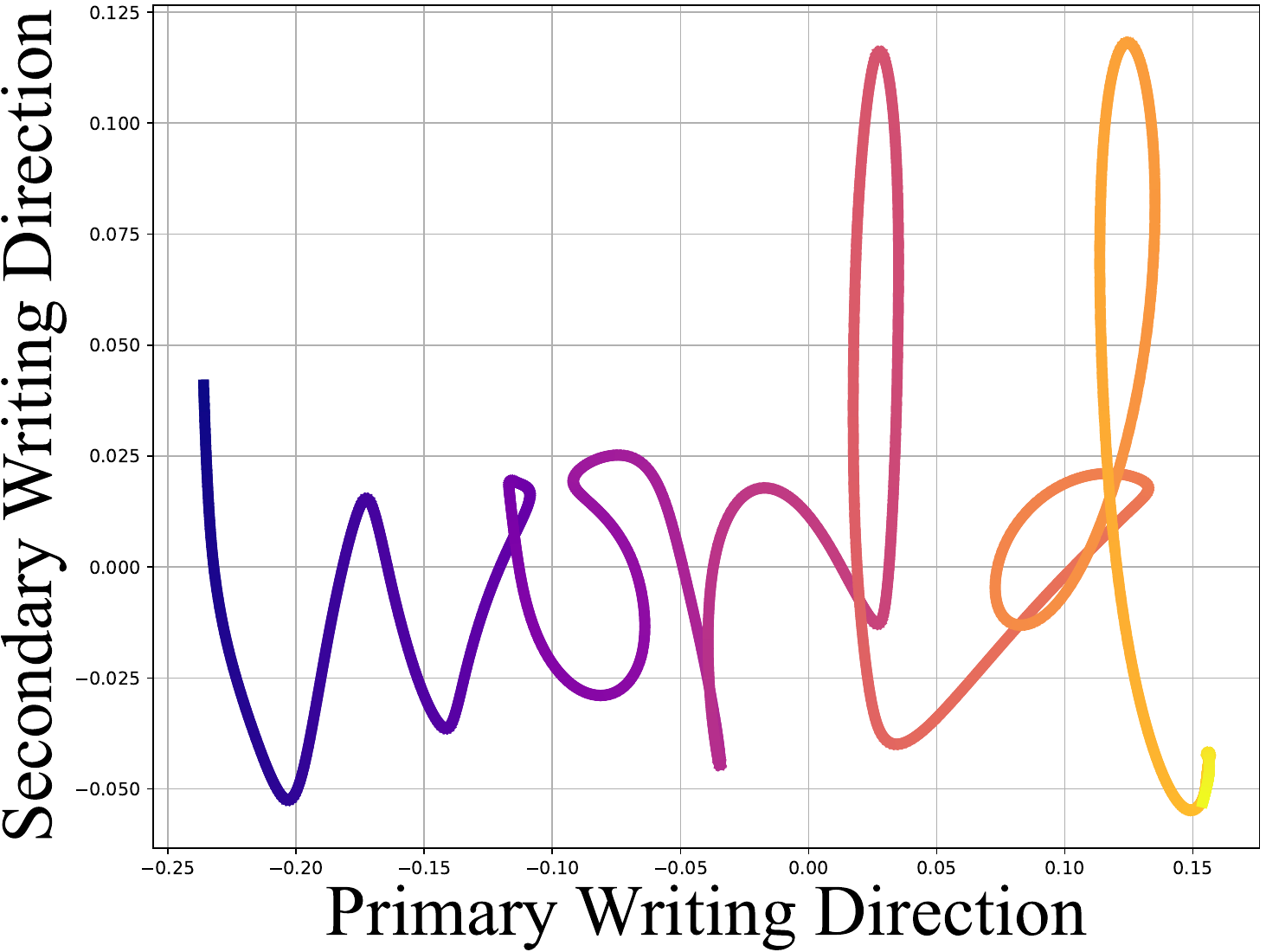}
  		\label{fig: world PCA}}
  	\end{minipage}
        \vspace{-2mm}
\caption{Writing trajectory images after direction-corrected PCA.}
\vspace{-3mm}
\label{fig: PCA}
\end{figure}

\section{Data Mapping from Spatial Trajectory to Word Image} \label{mapping}

In-air writing trajectories in space tend to be more arbitrary and irregular than those written on a plane. These trajectories can be influenced by hand movements, postures, or spatial orientations. Directly projecting them onto fixed coordinate axes may lead to significant distortion of the word's shape, posing challenges in maintaining the original structure. To address this issue, we employ PCA \cite{wold1987principal} to extract the principal directions of the spatial trajectory and transform it into an image, which helps retain the overall word shape effectively.

Specifically, we first convert the writing trajectories from spherical coordinates to Cartesian coordinates, where each trajectory point $\mathbf{x}_i = [x_i, y_i, z_i]^\top$ is represented as $ x = r \cdot \sin(\theta) \cdot \cos(\phi), y = r \cdot \cos(\theta) \cdot \cos(\phi), z = r \cdot \sin(\phi)$. Given the centralized handwriting trajectory sequence $\mathbf{X} = [\mathbf{x}_1, \mathbf{x}_2, \dots, \mathbf{x}_n]^\top \in \mathbb{R}^{n \times 3}$, we compute the covariance matrix $\mathbf{C}$ and perform eigenvalue decomposition as:
\begin{equation}
    \mathbf{C} \mathbf{V} = \mathbf{V} \mathbf{\Lambda}, \quad \mathbf{\Lambda} = \text{diag}(\lambda_1, \lambda_2, \lambda_3), \quad \lambda_1 \geq \lambda_2 \geq \lambda_3,
\end{equation}
where $\mathbf{\Lambda}$ is a diagonal matrix containing the eigenvalues, and $\mathbf{V} = [\mathbf{v}_1, \mathbf{v}_2, \mathbf{v}_3]$ is the matrix of eigenvectors. In normal handwriting, the principal directions of the trajectory are usually related to writing habits, with the primary direction representing horizontal extension (left-to-right writing) and the secondary direction reflecting vertical variations. Thus, we select the eigenvectors $\mathbf{v}_1$ and $\mathbf{v}_2$ corresponding to the two largest eigenvalues $\lambda_1$ and $\lambda_2$ to construct a 2D projection basis:
\begin{equation}
    \mathbf{P} = [\mathbf{v}_1, \mathbf{v}_2] \in \mathbb{R}^{3 \times 2}.
\end{equation}

These two principal writing directions precisely capture the characteristics of writing. Specifically, the largest eigenvalue corresponds to the primary horizontal writing trend, while the second largest eigenvalue corresponds to the secondary vertical variation trend. The trajectory points projected from space to plane can then be expressed as:
\begin{equation}
    \mathbf{X}_{2D} = \mathbf{X} \mathbf{P}, \quad \mathbf{X}_{2D} \in \mathbb{R}^{n \times 2},
\end{equation}
where $n$ represents the total number of trajectory points, and $\mathbf{X}_{2D}$ represents the generated word image. 

However, PCA determines the principal directions based on the statistical variance of spatial data, neglecting the sequential or temporal nature of handwriting trajectories. This limitation may result in dimensionality reduction results that deviate from typical writing habits, such as left-to-right progression, potentially resulting in horizontally or vertically mirrored trajectories. These distortions can impede the recognition of handwriting trajectories. In practical handwriting scenarios, words are typically written starting from the top-left corner and progressing towards the bottom-right corner. Based on this observation, we propose a post-processing adjustment method to correct the orientation of PCA results, ensuring that the writing direction aligns with common conventions and eliminating potential mirroring distortions. This adjustment process involves analyzing the starting and ending points of the PCA-reduced trajectory to determine the overall writing direction. Specifically, we compute the horizontal and vertical components of the trajectory's direction vector. If the horizontal component indicates a right-to-left progression, the trajectory is flipped horizontally to achieve left-to-right alignment. Similarly, if the vertical component indicates an upward trend, the trajectory is flipped vertically to conform to the top-to-bottom writing habit. Examples of gradient color visualization images are shown in Fig. \ref{fig: PCA}.

The direction-corrected PCA-based data mapping algorithm reconstructs the overall word shape and eliminates distortions caused by irregular spatial writing directions, such as slanted writing. It ensures that the generated trajectory follows logical writing habits, e.g., left-to-right progression, and aligns with visual conventions. By rectifying the trajectory and converting it into an image, we mitigate the impact of writing order on word recognition. This approach circumvents the obstacles that time-series-based techniques like LSTM frequently encounter.

\begin{figure}
	\centering
	\begin{minipage}[t]{\textwidth}
		\centering
  		\subfigure[Living room scene (4.2m $\times$ 3m).]{
  		\includegraphics[width=0.45\linewidth]{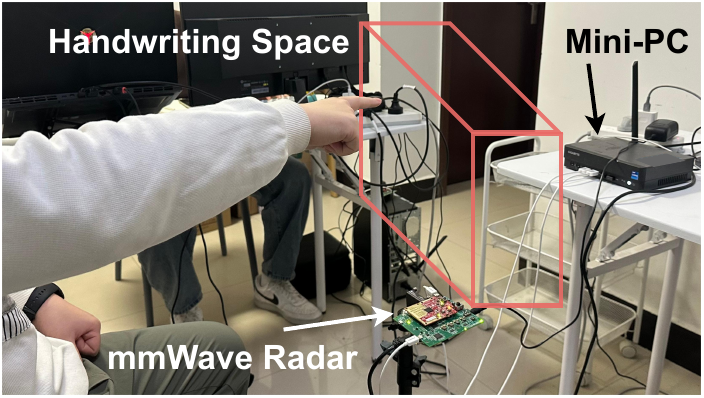}
  		\label{fig: living room}}
  		\subfigure[Bedroom scene (2.4m $\times$ 2.4m).]{
  		\includegraphics[width=0.45\linewidth]{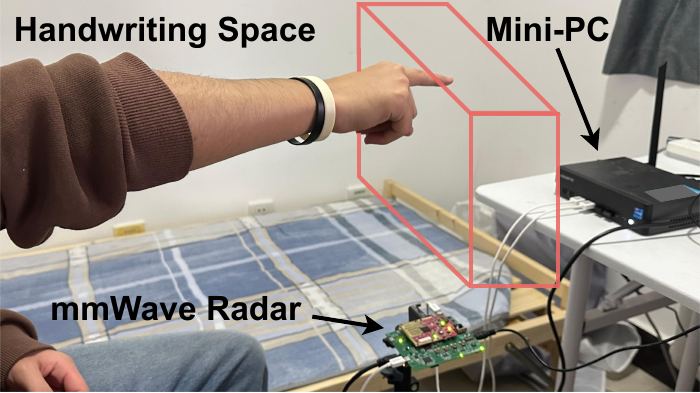}
  		\label{fig: bedroom}}
  	\end{minipage}
\caption{Data collection in home scenes.}
\label{fig: environment}
\end{figure}

\section{Evaluation} \label{evaluation}

To validate the performance of $m^2VLMs$, we develop a prototype in-air handwriting recognition system named \textit{mmPencil}. Specifically, a commercial mmWave radar is used to collect sensing data samples of 3D mmWave handwriting words, and a series of experiments are conducted to evaluate the accuracy and robustness of \textit{mmPencil} across a range of conditions, including different users, varying word lengths (i.e., number of letters), as well as diverse writing ranges, angles, speeds, sizes, directions and environmental factors.

\subsection{Experiment Setup}

\subsubsection{Hardware Implementation} 
To implement \textit{mmPencil}, we use the IWR6843ISK mmWave radar, which can transmit FM continuous wave signals in the frequency range of 60 to 64 GHz, and the DCA1000EVM for data collection. It is equipped with 3 transmitting antennas (one of which is arranged at different heights) and 4 receiving antennas, with a peak gain of 7 dBi, and can provide a horizontal FoV of $120^{\circ}$ and a elevation FoV of $60^{\circ}$. We set the frequency slope to 66.011 MHz/$\mu$s, and set the sampling rate to 2000 ksps. At the same time, each frame consists of one chirp with 108 ADC samples per chirp, resulting in a range resolution of 4.2 cm. In addition, the mmWave data collection module is deployed on a mini-PC running Windows 10, with a CPU frequency of 2.8 GHz. We use a GPU server equipped with 8 A100 GPUs (each with 40 GB of memory) for fine-tuning the VLMs.

\subsubsection{Dataset Collection}
\label{sec: dataset collection}

During the dataset collection process, the radar is stationary and placed facing upward. Users write at arbitrary positions within a $100cm \times 50cm \times 30cm$ space above the radar (as shown in Fig. \ref{fig: environment}). Beyond the requirement that the writing action takes place within the radar's field of view, there are no restrictions imposed on the writing position, order, or direction. In other words, users retain the freedom to adhere to their individual writing styles and habits without any limitations. The experiments involve 200 different words, each consisting of 2 to 9 letters, as shown in Table \ref{table: words}. These words are the most frequently occurring words in the IAM handwriting dataset \cite{marti2002iam}, which is one of the most well-known visual handwriting databases. Additionally, the duration of each writing action is related to the length of the word, and each letter being written within approximately 2 to 3 seconds. 

\begin{table}[t]
\caption{200 Categories of Words Contained in the \textit{mmPencil} Dataset.}
\label{table: words}
\small
\centering
\begin{tabular}{|c|c|c|c|c|c|c|c|c|c|}
\hline
about & action & again & all & almost & also & am & among & an & and \\ \hline
any & around & as & asked & at & away & back & be & because & become \\ \hline
been & before & being & believed & best & better & both & but & by & called \\ \hline
case & cent & change & come & could & course & day & did & do & does \\ \hline
down & early & end & enough & even & every & face & fact & few & film \\ \hline
find & first & for & found & four & from & full & get & give & given \\ \hline
go & god & going & good & great & had & half & hand & has & have \\ \hline
having & he & head & health & heard & heart & help & her & here & high \\ \hline
him & his & home & house & how & in & into & is & it & its \\ \hline
just & know & known & labour & last & later & lead & less & like & little \\ \hline
long & look & looked & make & man & many & may & me & men & might \\ \hline
most & much & my & national & need & never & new & next & no & not \\ \hline
number & of & old & on & once & one & only & or & our & out \\ \hline
own & party & people & per & place & play & point & policy & power & put \\ \hline
quite & really & said & say & see & seems & seen & she & should & shown \\ \hline
since & sir & small & so & some & something & still & strong & table & than \\ \hline
that & the & their & them & there & these & think & three & to & today \\ \hline
told & too & town & two & under & up & us & use & very & was \\ \hline
way & we & well & were & what & when & where & which & while & who \\ \hline
whom & why & will & world & wrote & years & yesterday & you & young & your \\ \hline
\end{tabular}
\end{table}

\begin{figure}
	\centering
	\begin{minipage}[t]{\textwidth}
		\centering
  	\subfigure[User 1.]{
  		\includegraphics[width=0.15\linewidth]{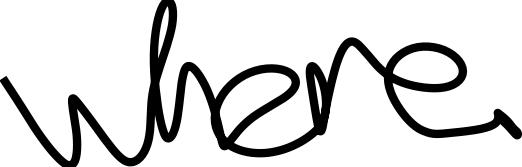}
  		\label{fig: GYF}}
        \subfigure[User 2.]{
  		\includegraphics[width=0.15\linewidth]{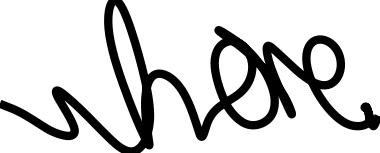}
  		\label{fig: QQ}}
        \subfigure[User 3.]{
  		\includegraphics[width=0.15\linewidth]{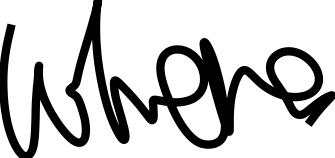}
  		\label{fig: GQW}}
        \subfigure[User 4.]{
  		\includegraphics[width=0.15\linewidth]{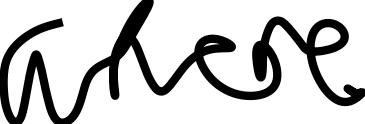}
  		\label{fig: SWC}}
        \subfigure[User 5.]{
  		\includegraphics[width=0.15\linewidth]{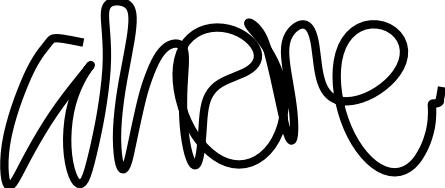}
  		\label{fig: ZQ}}
        \subfigure[User 6.]{
  		\includegraphics[width=0.15\linewidth]{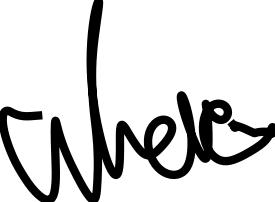}
  		\label{fig: XYD}}
        \subfigure[User 7.]{
  		\includegraphics[width=0.15\linewidth]{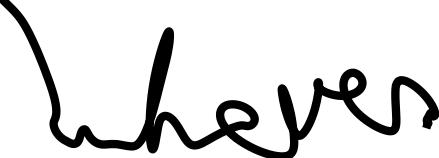}
  		\label{fig: ZHD}}
        \subfigure[User 8.]{
  		\includegraphics[width=0.15\linewidth]{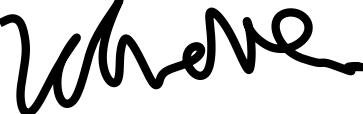}
  		\label{fig: CYF}}
        \subfigure[User 9.]{
  		\includegraphics[width=0.15\linewidth]{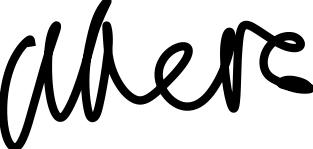}
  		\label{fig: LFJ}}
        \subfigure[User 10.]{
  		\includegraphics[width=0.15\linewidth]{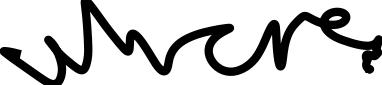}
  		\label{fig: ZHY}}
        \subfigure[User 11.]{
  		\includegraphics[width=0.15\linewidth]{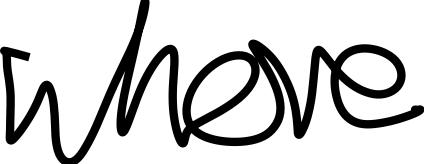}
  		\label{fig: MDL}}
        \subfigure[User 12.]{
  		\includegraphics[width=0.15\linewidth]{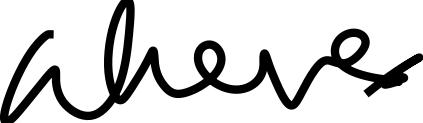}
  		\label{fig: LYQ}}
  	\end{minipage}
\caption{In-air handwriting style of all users (case word "where").}
\label{fig: writing style}
\end{figure}

Drawing on previous studies experience \cite{10.1145/2858036.2858498}, we recruit 12 users, including 9 non-authors, whose detailed information is summarized in Table \ref{table: volunteer}. In addition, Fig. \ref{fig: writing style} illustrates representative examples of each volunteer’s in-air handwriting style. The collected dataset consists of three complementary parts, each designed to address different research focuses, with the first and second parts covering partially overlapping content, as follows:
\begin{itemize}
\item \textbf{\textit{200-Word 4-User Dataset.}} The first part targets a broader range of word categories and is contributed by 4 users, each collecting samples for 200 distinct words. Specifically, users 1 and 2 collect 10 samples per word, whereas users 3 and 4 collect 2 samples per word, yielding a total of 4,800 samples. 

\item \textbf{\textit{50-Word 12-User Dataset.}} The second part of the dataset emphasizes the diversity of handwriting styles across all 12 users. Specifically, we first extract data corresponding to 50 word categories (comprising 2 to 9 letters each) from 4 users in the first part of the dataset. Subsequently, we incorporate additional data collected from the remaining 8 users, each contributing samples for the same 50 word categories, 10 numerical categories (0-9), and 2 symbol categories ($\rightarrow$ and $+$), with 4 samples per category. In total, 3184 samples are generated, among which 1984 are non-duplicate samples.

\item \textbf{\textit{22-Writing-Scenario 1-User Dataset.}} The third part focuses on diverse handwriting scenarios, where User 1 performs handwriting under 22 different scenarios, including variations in writing ranges, angles, speeds, environments, sizes, directions and user movements. For each scenario, 8 word categories are included with 5 samples per word, amounting to 880 samples.
\end{itemize}
Overall, We collect a total of 7,664 data samples of 12 different users across various scenarios, resulting in a dataset of about 31.66 GB.

\begin{table}[tb]
  \caption{Information of Participant Users and Collected Data.}
  \label{table: volunteer}
  \small
  \begin{tabular}{ccccc}
    \toprule
    User Number & Age & Occupation & Data Scope & Total Data Volume\\
    \midrule
    1 & 28 & Ph.D. Student & 200 words & 2880 \\
    2 & 23 & Engineering Intern & 200 words & 2000 \\
    3 & 24 & Master's Student & 200 words & 400 \\
    4 & 28 & Assistant Engineer & 200 words & 400 \\
    5 & 24 & Engineering Intern & 50 words \& 10 numbers \& 2 symbols & 248 \\
    6 & 24 & Ph.D. Student & 50 words \& 10 numbers \& 2 symbols & 248 \\
    7 & 22 & Master's Student & 50 words \& 10 numbers \& 2 symbols & 248 \\
    8 & 22 & Master's Student & 50 words \& 10 numbers \& 2 symbols & 248 \\
    9 & 24 & Engineering Intern & 50 words \& 10 numbers \& 2 symbols & 248 \\
    10 & 24 & Engineering Intern & 50 words \& 10 numbers \& 2 symbols & 248 \\
    11 & 24 & Engineering Intern & 50 words \& 10 numbers \& 2 symbols & 248 \\
    12 & 24 & Engineering Intern & 50 words \& 10 numbers \& 2 symbols & 248 \\
    \bottomrule
\end{tabular}
\vspace{-5mm}
\end{table}

\subsubsection{Fine Tuning Setup for VLMs}
To demonstrate the versatility of the proposed architecture in supporting various VLMs, we select three open-source VLMs: GOT-OCR-0.5B \cite{wei2024general}, Qwen2-VL-7B \cite{wang2024qwen2} and LLaVA-NeXT-7B \cite{liu2024llavanext}. Prior to evaluation, we fine-tune Qwen2-VL-7B and LLaVA-NeXT-7B with Llama Factory \cite{zheng2024llamafactory}, and fine-tune GOT-OCR-0.5B with Swift \cite{zhao2025swift}. Specifically, we apply the Low-Rank Adaptation (LoRA) \cite{hu2022lora} technique to fine-tune three pretrained models. The rank and scaling factor of LoRA are set to $8$ and $16$, and LoRA is applied to all linear layers of the models to achieve more comprehensive parameter optimization. During training, we use a learning rate of $5 \times 10^{-5}$ and train the models for $10$ epochs.

These refined VLMs exhibit enhanced capability in processing RF handwriting data, particularly in scenarios involving variations in handwriting styles or environmental conditions. Notably, the instruction fine-tuning approach does not require large amounts of data for training from scratch, which is often unattainable in the RF sensing domain due to data scarcity and labeling complexities. Instead, a small amount of data is sufficient to significantly improve the performance of handwriting word recognition.

\subsection{Accuracy Evaluation} \label{performance}
To comprehensively evaluate the recognition performance of the system, we conduct both non-cross-user (i.e., training and testing on the same users) and cross-user (i.e., training and testing on different users) evaluations on the \textit{200-Word 4-User Dataset} and \textit{50-Word 12-User Dataset}. Subsequently, we systematically evaluate the system’s cross-user recognition performance on the \textit{50-Word 12-User Dataset} under different numbers of seen users during training. Finally, we conduct ablation studies on the $mmPencil$ system.

\subsubsection{Performance Comparison on the 200-Word 4-User Dataset}
\label{sec: 200word}

\begin{figure}
	\centering
	\begin{minipage}[ht]{\textwidth}
		\centering
  		\subfigure[Results based on the 200-Word 4-User Dataset.]{
  		\includegraphics[width=0.45\linewidth]{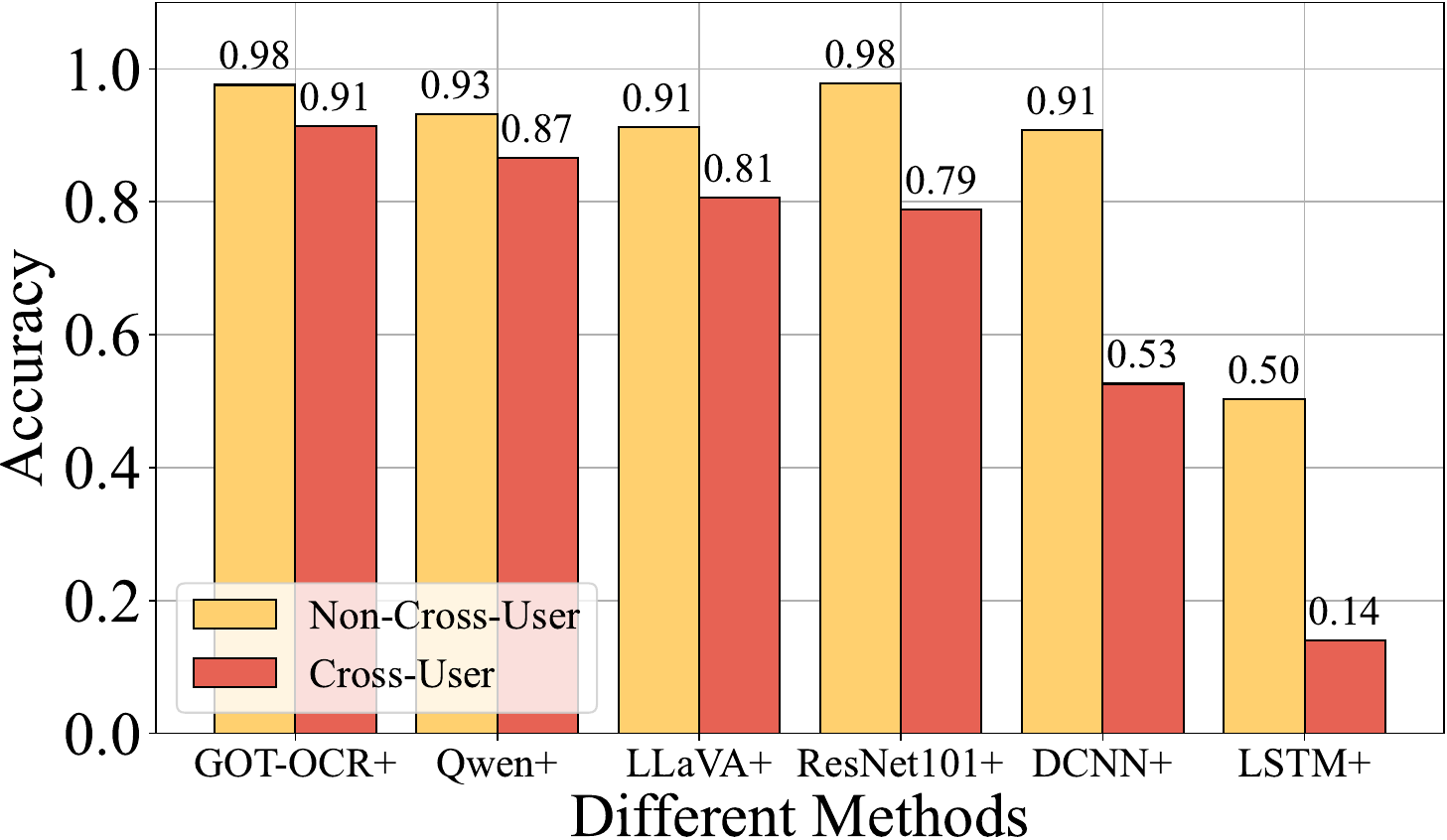}
  		\label{fig: 4v_200c}}\hspace{5mm}
  		\subfigure[Results based on the 50-Word 12-User Dataset.]{
  		\includegraphics[width=0.45\linewidth]{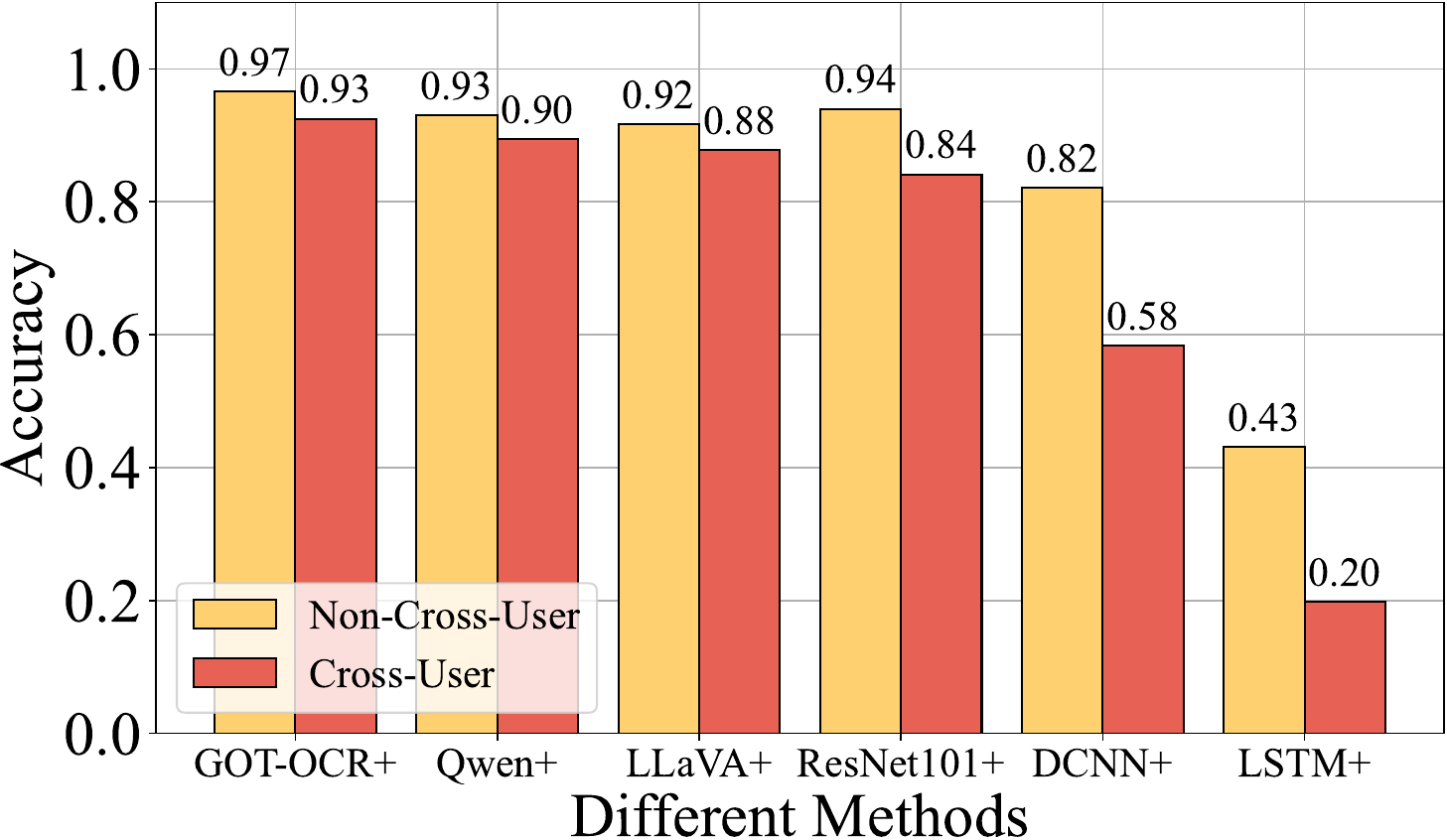}
  		\label{fig: 12v_50c}}
  	\end{minipage}
        \vspace{-3mm}
\caption{Performance comparison with baselines ("+" represents the trained or fine-tuned model).}
\end{figure}

\begin{figure}
	\centering
	\begin{minipage}[th]{\textwidth}
		\centering
  		\subfigure[User 2.]{
  		\includegraphics[width=0.25\linewidth]{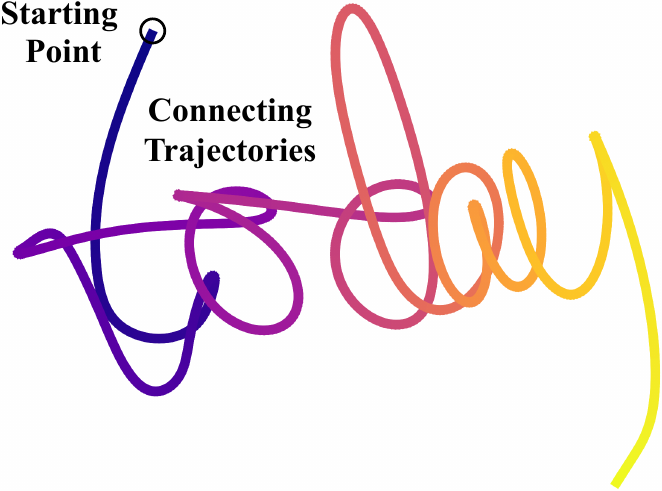}
  		\label{fig: User 2}}\hspace{10mm}
  		\subfigure[User 3.]{
  		\includegraphics[width=0.25\linewidth]{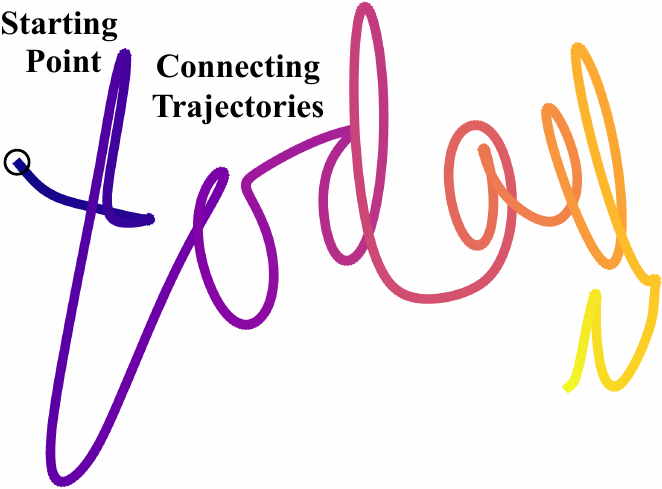}
  		\label{fig: User 3}}\hspace{10mm}
            \subfigure[User 4.]{
  		\includegraphics[width=0.25\linewidth]{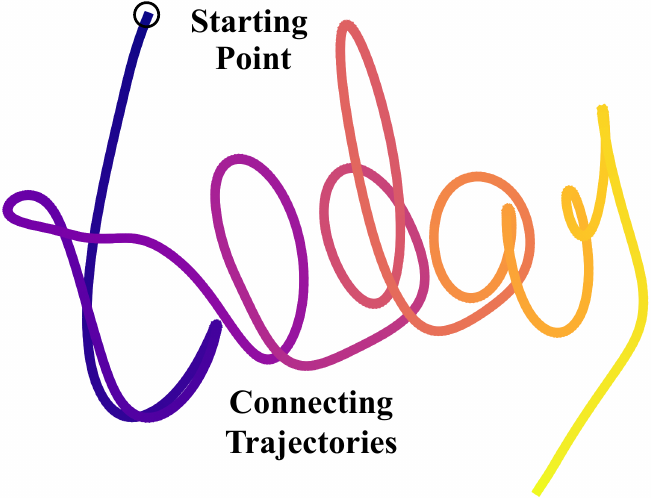}
  		\label{fig: User 4}}
  	\end{minipage}
\caption{Mapped images of the word "today" for different users.}
\label{fig: writing styles}
\vspace{-2mm}
\end{figure}

To highlight the advantages of the \textit{mmPencil} system, we compare its performance with a set of state-of-the-art baselines. Specifically, we reproduce two baseline methods in the field of mmWave handwriting recognition, which are based on DCNN \cite{arsalan2019character} and LSTM \cite{chen2023mmhtsr}, respectively. However, as these methods are not designed for word-level 3D handwriting recognition, we adopt our proposed trajectory reconstruction algorithm to achieve a fair comparison. Since the relevant prior work does not release its dataset, we can only reproduce the DCNN model architecture based on the corresponding descriptions in the paper, but cannot obtain the pre-trained model parameters. Therefore, we train the DCNN model from scratch using the data collected in this study with randomly initialized weights. To further enhance the fairness of the comparison, we also introduce a pre-trained \textit{ResNet-101} \cite{he2016deep} model as an additional baseline.

For the LSTM-based method, we do not simply transform the data into 2D images for recognition. Instead, we strictly follow the original methodology by directly training and evaluating the model on the 3D spatial trajectory sequences generated by the $m^2VLMs$ architecture, thus preserving the original setting and ensuring the validity of the comparison. During training, the baseline models typically achieve stable convergence within 100 to 200 epochs.

We first evaluate the performance of the \textit{mmPencil} system based on 6 different methods (including 3 VLMs-based methods: \textit{mmPencil-GOT}, \textit{mmPencil-Qwen}, and \textit{mmPencil-LLaVA}, as well as 3 baseline methods: \textit{mmPencil-ResNet}, \textit{mmPencil-DCNN}, and \textit{mmPencil-LSTM}) on the \textit{200-Word 4-User Dataset}. Specifically, in the \textit{non-cross-user} setting, data samples are randomly partitioned into a 19:5 ratio for training and testing, while in the \textit{cross-user} setting, the training dataset comprises samples from users 1 and 2, and the testing dataset include samples from users 3 and 4. 

As shown in Fig. \ref{fig: 4v_200c}, due to significant variations in handwriting styles, all systems achieve higher performance in the non-cross-user setting than in the cross-user setting. In the non-cross-user experiments, benefiting from the high-quality writing trajectories reconstructed by the proposed algorithm, all methods except the one based on LSTM achieve recognition accuracies above 90\%. Among them, \textit{mmPencil-GOT} and \textit{mmPencil-ResNet} achieve the highest accuracies of 97.60\% and 97.80\%, respectively. In comparison, \textit{mmPencil-Qwen}, \textit{mmPencil-LLaVA}, and \textit{mmPencil-DCNN} show slightly lower but still good performances, achieving accuracies of 93.20\%, 91.30\%, and 90.80\%, respectively. In the cross-user experiments, methods based on VLMs demonstrate superior generalization ability, even when trained with data from only 2 users, and the corresponding accuracies are 91.37\%, 86.62\%, and 80.63\%, respectively. In contrast, both \textit{mmPencil-ResNet} and \textit{mmPencil-DCNN} remain highly sensitive to variations in writing styles, leading to notable performance drops, with accuracies of 78.87\% and 52.62\%, respectively.

It is worth noting that, due to inevitable cursive connections between letters during in-air word writing, word-level recognition becomes considerably more challenging than traditional isolated letter recognition. Moreover, variations in stroke orders among different users further degrade the performance of LSTM-based methods in the cross-user scenario. As shown in Fig. \ref{fig: writing styles}, taking the word "today" as an example, User 2 and User 3 tend to connect letters (e.g., "t-o-d") above the baseline, whereas User 4 prefers to connecting them below the baseline. Additionally, User 2 and User 4 typically write the letter "t" from top to bottom, whereas User 3 starts writing from left to right. These style discrepancies ultimately limit the performance of the \textit{mmPencil-LSTM} model, which achieves only 50.30\% and 14.00\% accuracy in the non-cross-user and cross-user settings, respectively.

In conclusion, the experimental results demonstrate that the proposed $m^2VLMs$ architecture effectively bridges the gap between RF sensing data and VLMs, enabling \textit{mmPencil} to surpass existing solutions by a significant margin in word-level handwriting recognition tasks.

\subsubsection{Performance Comparison on the 50-Word 12-User Dataset}
\label{sec: 12-user}

To further evaluate the system’s generalization ability to diverse users, we conduct experiments on the \textit{50-Word 12-User Dataset} using the same 6 different \textit{mmPencil} configurations as \S\ref{sec: 200word}. The overall recognition performance is shown in Fig. \ref{fig: 12v_50c}. Specifically, in the \textit{non-cross-user} setting, samples 
of each user are randomly split into a 44:12 ratio for training and testing. In contrast, in the \textit{cross-user} setting uses samples of 8 randomly selected participants are used for training, and samples of the remaining 4 users are used for testing.

Compared to the results obtained based on the \textit{200-Word 4-User Dataset} shown in Fig. \ref{fig: 4v_200c}, the recognition accuracy of all models (except \textit{mmPencil-LLaVA}) decreases under the non-cross-user setting. Among them, the performance drop is particularly pronounced for lightweight baseline models: \textit{mmPencil-ResNet}, \textit{mmPencil-DCNN}, and \textit{mmPencil-LSTM} decrease to 94.00\%, 82.17\%, and 43.17\%, respectively. In comparison, VLM-based models are less affected: \textit{mmPencil-GOT} and \textit{mmPencil-Qwen} achieve accuracies of 96.67\% and 93.00\%, while \textit{mmPencil-LLaVA} even shows a slight improvement to 91.67\%. Conversely, in the cross-user setting, all models exhibit improved performance due to the benefit of more diverse data samples. VLMs-based methods consistently demonstrate superior generalization capabilities, and the recognition accuracies of \textit{mmPencil-GOT}, \textit{mmPencil-Qwen}, and \textit{mmPencil-LLaVA} are 92.50\%, 89.50\%, and 87.86\%, respectively. Meanwhile, the performance of \textit{mmPencil-ResNet}, \textit{mmPencil-DCNN}, and \textit{mmPencil-LSTM} also improved to 84.13\%, 58.33\%, and 19.83\%, respectively.

Overall, as the number of users in the training data increases, both VLMs-based and lightweight \textit{mmPencil} models experience enhanced generalization. Nevertheless, VLMs-based models consistently maintain higher recognition accuracy and exhibit significantly greater robustness to handwriting style variations. These findings highlight the crucial advantage of using large pre-trained models for in-air handwriting recognition, offering stronger cross-user generalization and greater potential for real-world deployment.

\begin{figure}
	\centering
	\begin{minipage}[t]{0.9\textwidth}
		\centering
  		\includegraphics[width=\linewidth]
            {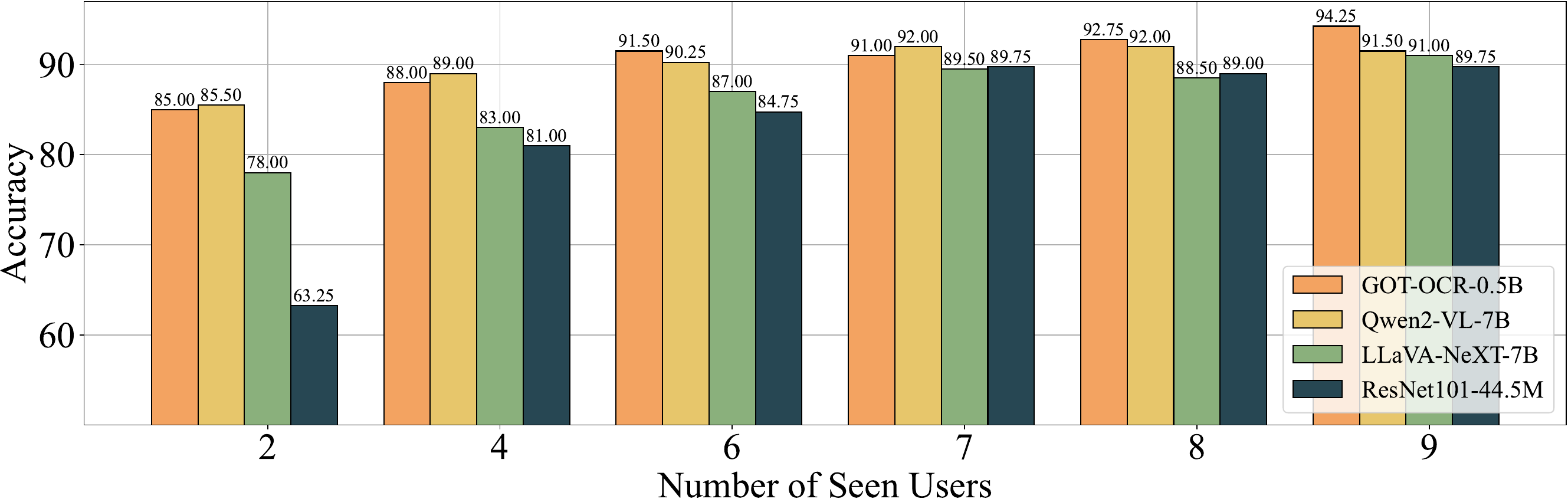}
            \caption{Recognition accuracy of unseen users.}
  		\label{fig: unseen users}
        \end{minipage}
        \vspace{-5mm}
\end{figure}

\subsubsection{Cross-User Recognition Performance under Varying Numbers of Seen Users}

Furthermore, We systematically evaluate the cross-user recognition performance of the \textit{mmPencil} system under different numbers of seen users on the \textit{50-Word 12-User Dataset}. With data samples from 3 randomly selected users as the test set, we vary the number of seen users during training (2, 4, 6, 7, 8, and 9) and evaluate the performance of 4 different models: \textit{mmPencil-GOT}, \textit{mmPencil-Qwen}, \textit{mmPencil-LLaVA}, and \textit{mmPencil-ResNet}.

As shown in Fig. \ref{fig: unseen users}, all models exhibit consistent performance improvements as the number of seen users increases. Among them, \textit{mmPencil-GOT} achieves the best overall performance, with the accuracy steadily improving from 85.00\% with 2 seen users to 94.25\% for 9 seen users. Similarly, \textit{mmPencil-Qwen} demonstrates strong generalization capability, reaching an accuracy of 91.50\% when trained with the data of 9 users. In contrast, \textit{mmPencil-LLaVA} starts with a lower initial accuracy but shows substantial improvement, increasing from 78.00\% to 91.00\%. Similarly, the lightweight \textit{mmPencil-ResNet} model also exhibits a significant improvement, rising from 63.25\% to 89.75\% as the number of seen users grows.

Overall, the VLMs-based \textit{mmPencil} models, particularly \textit{mmPencil-GOT} and \textit{mmPencil-Qwen}, demonstrate stronger generalization capabilities even with a small number of seen users. Although lightweight models such as \textit{mmPencil-ResNet} also benefit significantly from more diverse training data, their performance remains notably lower than that of VLMs-based models when the number of seen users is limited, further highlighting the advantage of large models in complex cross-user scenarios.

\subsubsection{Ablation Studies}
\begin{figure}
	\centering
        \begin{minipage}[t]{0.315\textwidth}
		\centering
  		\includegraphics[width=\linewidth]
            {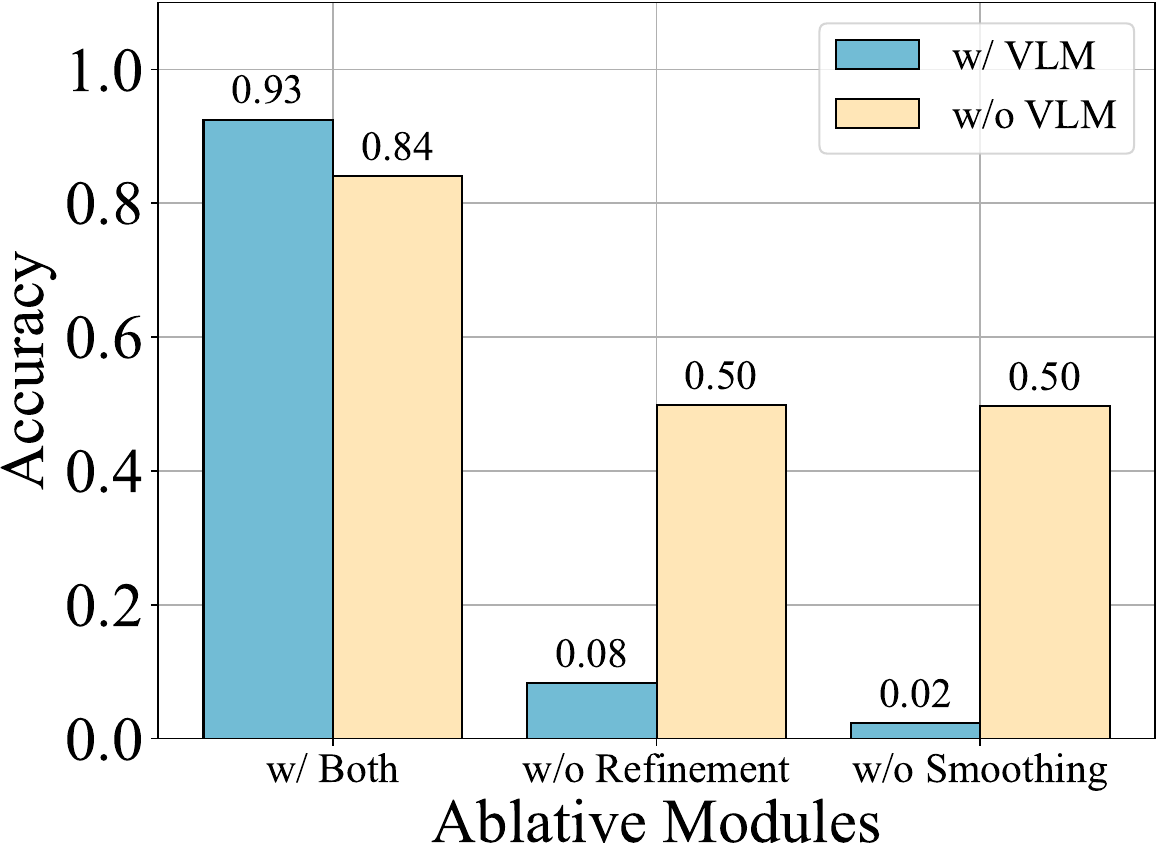}
            \caption{Recognition accuracy of ablation experiments.}
  		\label{fig: ablation}
        \end{minipage}\hspace{2mm}
        \begin{minipage}[t]{0.315\textwidth}
		\centering
  		\includegraphics[width=\linewidth]
            {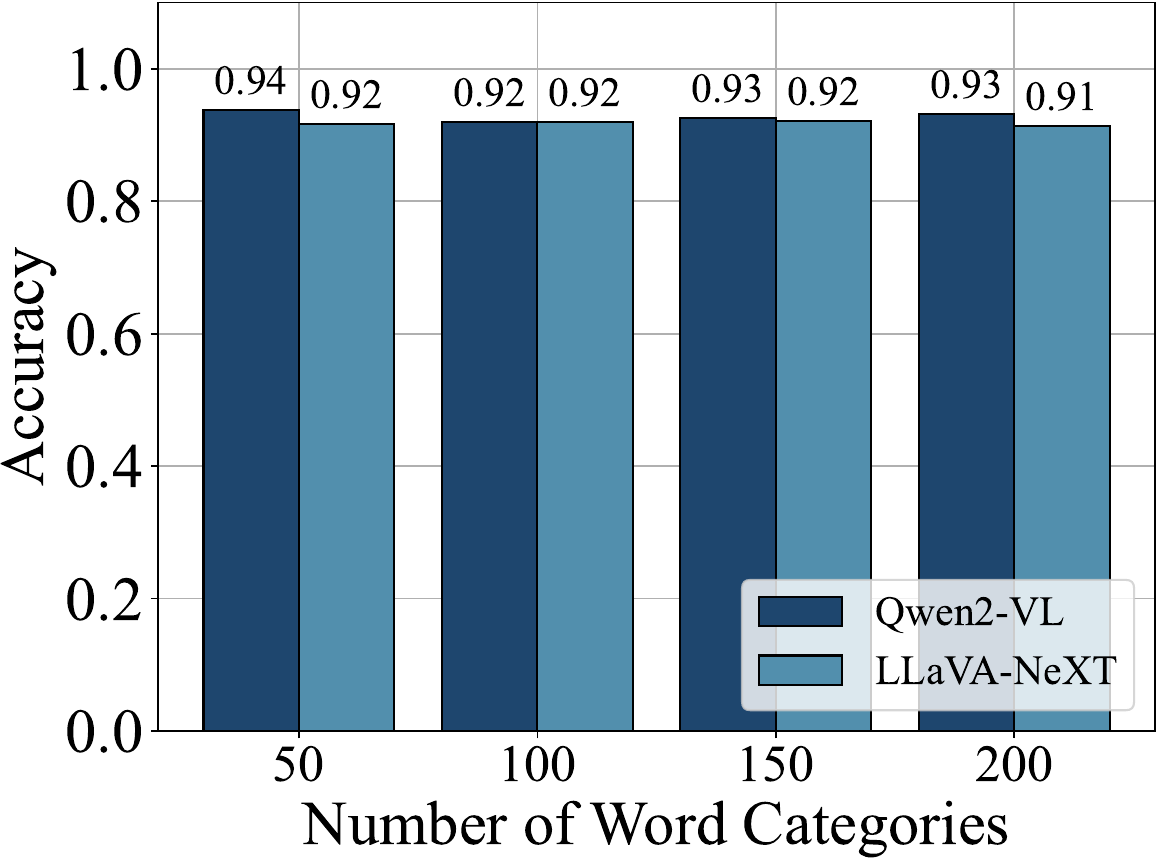}
            \caption{Recognition accuracy of experiments with different number of words.}
  		\label{fig: categories number}
        \end{minipage}\hspace{2mm}
	\begin{minipage}[t]{0.315\textwidth}
		\centering
  		\includegraphics[width=\linewidth]
            {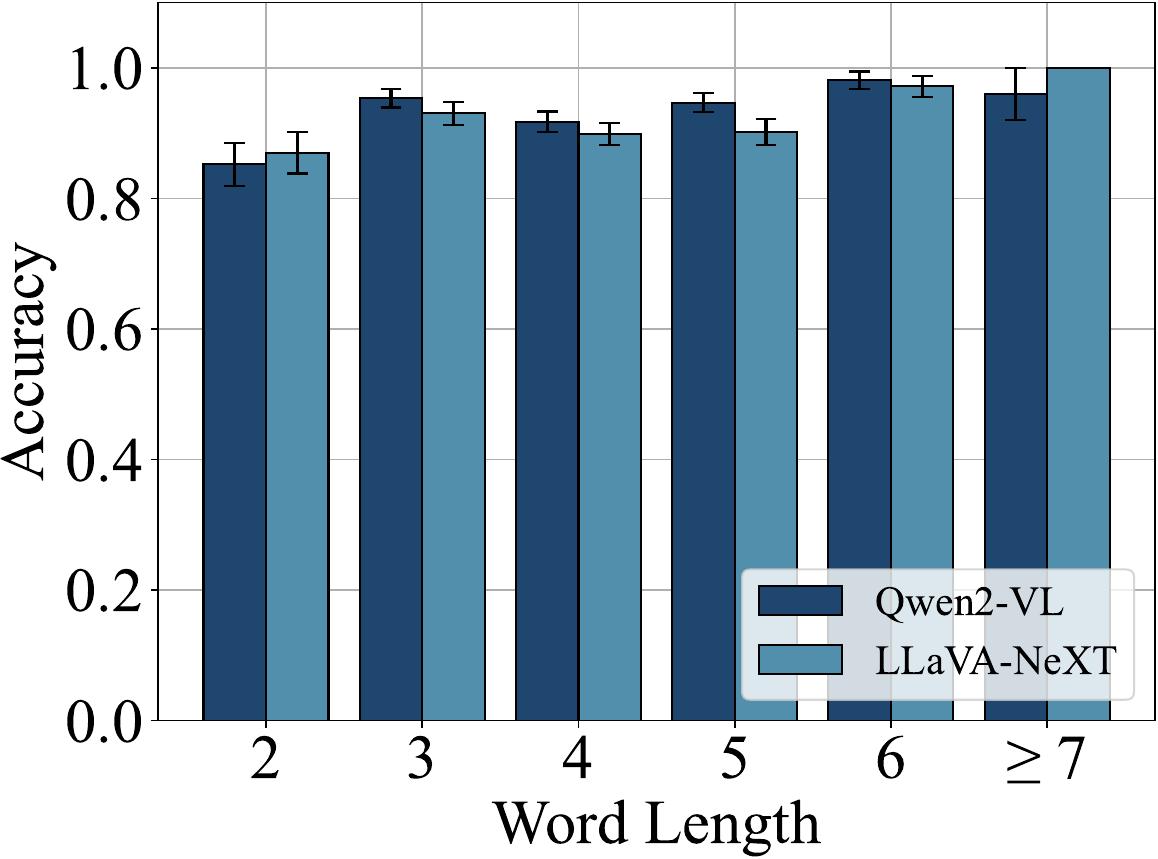}
            \caption{Recognition accuracy of experiments with different word lengths.}
  		\label{fig: length independence}
        \end{minipage}\hspace{2mm}
\end{figure}

Since the modules in the $m^2VLMs$ architecture are tightly connected in series, removing key components leads to significant performance degradation or even complete system failure. To evaluate the contribution of each module to the overall performance, we conduct ablation studies targeting three main components: Spectral Refinement, Trajectory Smoothing, and the Vision-Language Model. Specifically, we perform cross-user evaluations on the \textit{50-Word 12-User Dataset}, following the same training settings as described in \S\ref{sec: 12-user} In each ablation study, we remove one component: for Spectral Refinement, we eliminate the interpolation step used to enhance resolution; for Trajectory Smoothing, we remove the Kalman filter; and for the Visual-Language Model, we replace \textit{GOT-OCR} with a lightweight \textit{ResNet-101}. According to the experimental results shown in Fig. \ref{fig: ablation}, we fine that the complete $mmPencil$ system achieves a recognition accuracy of 92.50\%. As a comparison, when removing the Spectral Refinement module, the accuracy sharply drops to 8.33\%, and further drops to 2.33\% in case that the Trajectory Smoothing module is removed as well. These results highlight that both modules are indispensable for recovering high-quality handwriting trajectories and ensuring the final recognition performance. Although the impact of replacing \textit{GOT-OCR} with \textit{ResNet-101} is relatively small, the recognition accuracy still decreases to 84.13\%.

It is worth noting that if \textit{GOT-OCR} is replaced with \textit{ResNet-101}, the system's recognition accuracy remains as high as 49.83\% and 49.66\% in case that the Spectral Refinement or Trajectory Smoothing module is eliminated. This observation indicates that lightweight classifiers primarily rely on clustering simple image features. Thus, even when the trajectory is heavily distorted, they can still perform classification to some extent based on local patterns, without truly understanding the image content. In contrast, VLMs possess reasoning abilities and make recognition decisions by accurately understanding the handwritten content in the image. Therefore, when the trajectory is completely distorted, the reasoning process collapses, leading to recognition failure. However, as long as Spectral Refinement and Trajectory Smoothing modules effectively restore near-authentic handwriting trajectories, VLMs can fully unleash their understanding and reasoning capabilities, enabling not only accurate recognition of seen words but also unseen words.

\subsection{Robustness Evaluation}

We further evaluate the system’s robustness across different writing scenarios using the \textit{22-Writing-Scenario 1-User Dataset}, which covers a variety of writing ranges, angles, speeds, environments, sizes, directions, and user movements. Through these experiments, we demonstrate that the \textit{mmPencil} system can adapt to diverse and challenging conditions. To provide a more intuitive comparison of the impact of different writing scenarios, we employ a unified user-dependent model during the experiment. Specifically, the \textit{200-Word 4-User Dataset} is divided into training and testing sets with a 19:5 ratio. The training set is used to fine-tune two general-purpose vision-language models, namely Qwen2-VL and LLaVA-NeXT, which are subsequently applied across all robustness evaluations. For the testing data, we use the initially defined test split in \S\ref{sec: word-category-number independence} and \S\ref{sec: word-length independence}, while the following 7 robustness experiments are conducted using the \textit{22-Writing-Scenario 1-User Dataset}.

\subsubsection{Word-Category-Number Independence Evaluation}
\label{sec: word-category-number independence}

To avoid potential interference due to the reduced number of word categories in the newly collected supplementary dataset (i.e., from \textit{200-Word 4-User Dataset} to \textit{50-Word 12-User Dataset}), we conduct a word-category-number independence evaluation. Specifically, based on the \textit{200-Word 4-User Dataset}, we randomly sample subsets with 50, 100, and 150 categories, and train and evaluate different models separately, set of 200 categories serving as a baseline. As shown in Fig. \ref{fig: categories number}, for the \textit{mmPencil-Qwen} model, the recognition accuracies for 50, 100, 150, and 200 categories are 93.70\%, 92.00\%, 92.50\%, and 93.20\%, respectively. For the \textit{mmPencil-LLaVA} model, the corresponding accuracies are 91.70\%, 91.90\%, 92.10\%, and 91.30\%. The results indicate that the overall recognition performance fluctuates very slightly within a narrow range of approximately 2\% when the number of categories changes. This demonstrates that the developed \textit{mmPencil} system maintains strong stability in word recognition tasks across varying category scales. Even when the number of categories is significantly reduced (e.g., from 200 to 50), the system still achieves consistently high recognition performance. These findings further confirm that \textit{mmPencil} possesses excellent generalization capabilities for in-air handwritten word recognition, relying primarily on the spatial movement patterns of handwriting rather than the scale of category numbers.

\subsubsection{Word-Length Independence Evaluation}
\label{sec: word-length independence}

Results of \textit{mmPencil-Qwen} and \textit{mmPencil-LLaVA} show that the average accuracy for recognizing 200 categories of words in the test set reaches 93.20\% and 91.30\%, respectively. The performance for words of different lengths as shown in Fig. \ref{fig: length independence}. Contrary to traditional belief, we find that as the number of letters in a word increases, the accuracy does not decrease. Instead, the recognition rate of \textit{mmPencil-Qwen} gradually improves as the word length increases, increasing from 85.22\% (2 letters) to 95.35\% (3 letters), 91.75\% (4 letters), 94.67\% (5 letters), 98.10\% (6 letters), and reaching 96.00\% for words with 7 or more letters. \textit{mmPencil-LLaVA} also demonstrates relatively balanced performance across words of different lengths, with accuracies of 86.96\% for 2-letter words, 93.02\% for 3-letter words, 89.84\% for 4-letter words, 90.22\% for 5-letter words, 97.14\% for 6-letter words, and 100.00\% for words with 7 or more letters. This finding challenges the conventional notion that longer words are more difficult to recognize and are typically prone to more confusion.

We attribute this result to the pre-trained VLM's strong ability to understand contextual information, as well as the powerful ability of the trajectory reconstruction algorithm to restore the fine details of handwriting words. Longer words usually contain more context, which provides significant support for the model during the recognition process. In traditional methods, computer vision models typically rely on local features for recognition and struggle to handle complex letter combinations and contextual information. However, with the support of the large model, the additional clues provided by longer words help the model better understand the relationships between letters, improving accuracy. For example, for the word "about", even if there is an initial mis-prediction of some letters (e.g., "abaut"), the model can still correct the mistake and identify the correct word, aided by the instruction prompts and contextual reasoning. This ability to rely on context to correct initial judgments is something traditional systems cannot achieve.

To sum up, due to VLMs' proficiency in capturing contextual information within words, \textit{mmPencil} achieves remarkable performance, particularly in the recognition of longer words..

\begin{figure}
	\centering
        \begin{minipage}[t]{0.315\textwidth}
		\centering
  		\includegraphics[width=\linewidth]
            {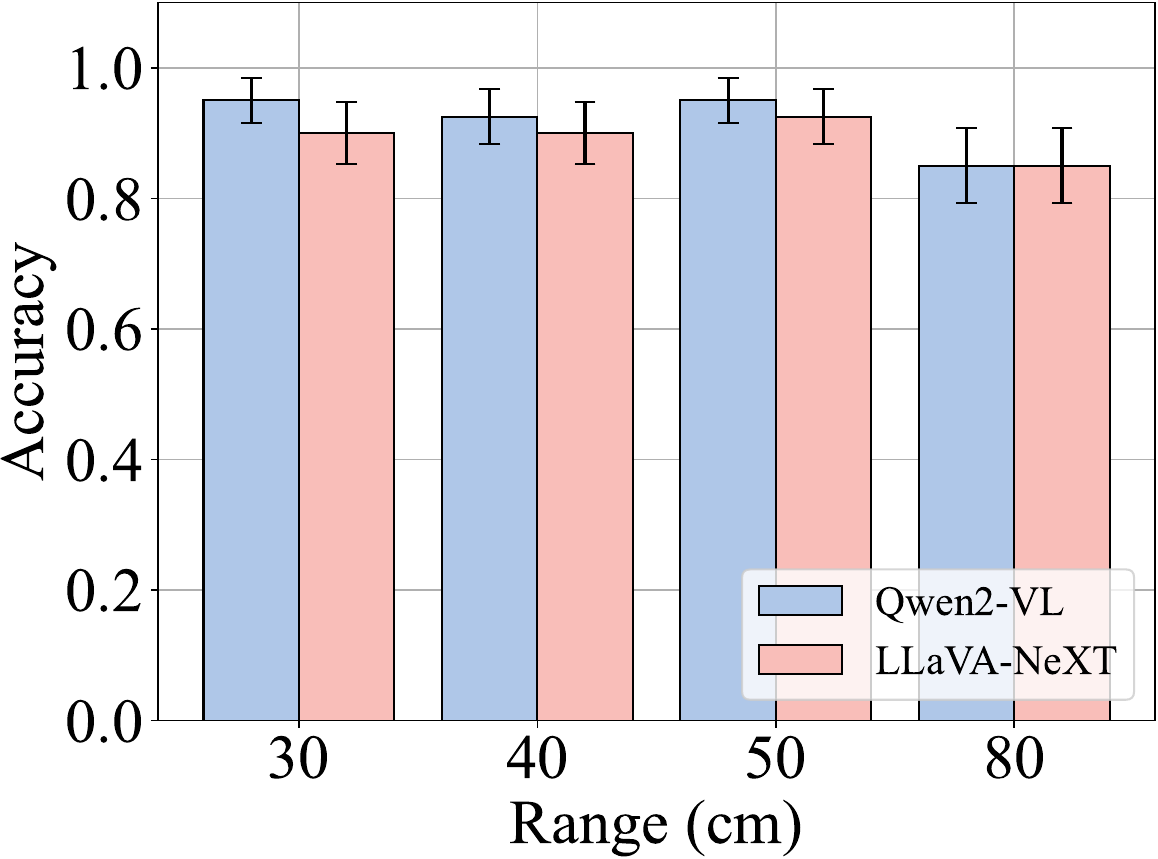}
            \caption{Recognition accuracy of experiments with different ranges.}
  		\label{fig: range independence}
        \end{minipage}\hspace{2mm}
        \begin{minipage}[t]{0.315\textwidth}
		\centering
  		\includegraphics[width=\linewidth]
            {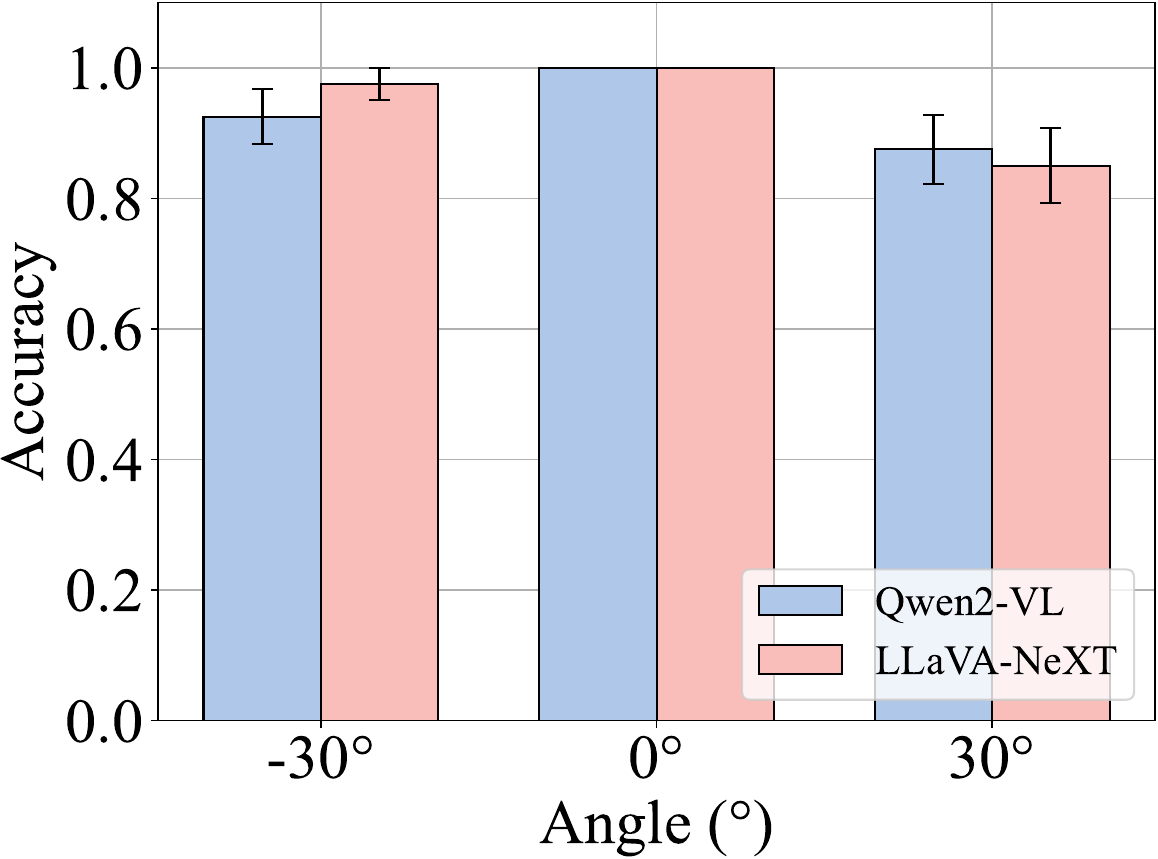}
            \caption{Recognition accuracy of experiments with different angles.}
  		\label{fig: angle independence}
        \end{minipage}\hspace{2mm}
        \begin{minipage}[t]{0.315\textwidth}
		\centering
  		\includegraphics[width=\linewidth]
            {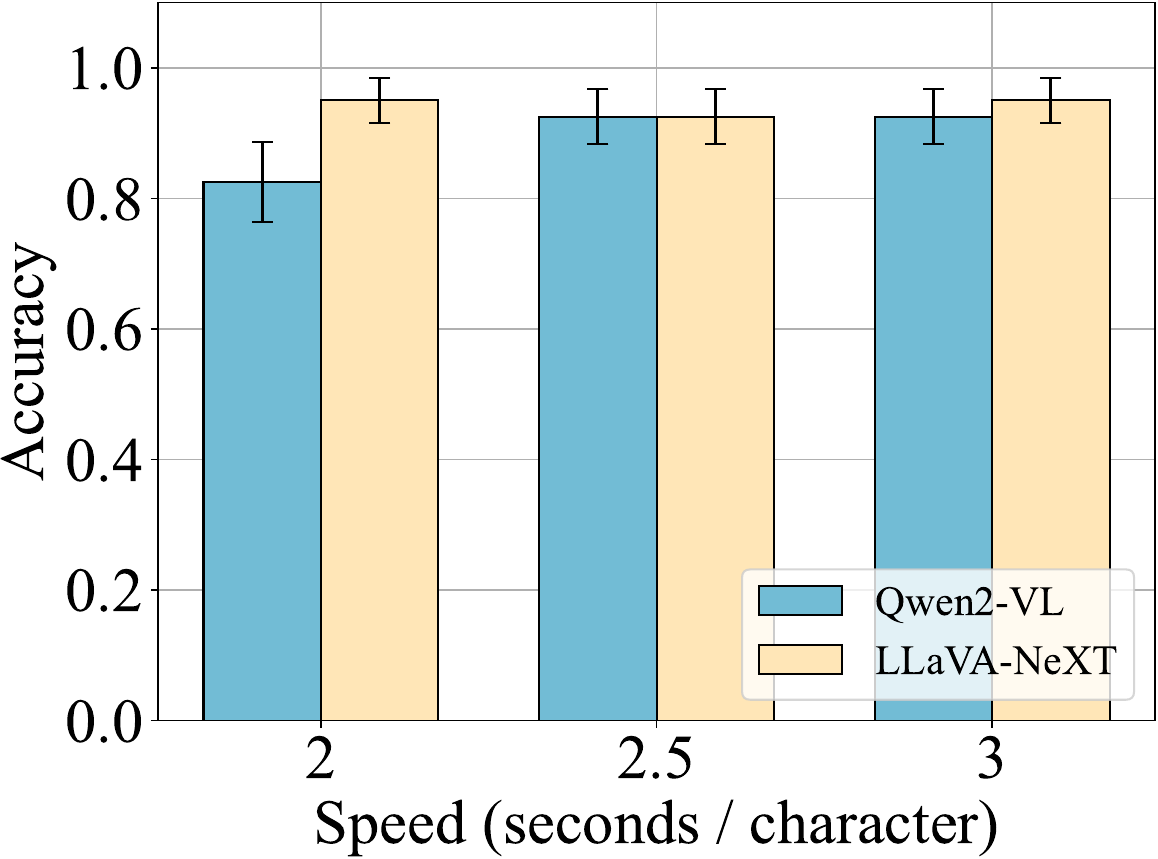}
            \caption{Recognition accuracy of experiments with different writing speeds.}
  		\label{fig: speed independence}
        \end{minipage}
\end{figure}

\subsubsection{Writing-Range Independence Evaluation}

To evaluate the system's performance at different writing ranges, we randomly select 8 types of words of 4 different lengths (including "and", "day", "been", "here", "seems", "where", "enough", and "really") and conduct experiments at close ranges (30 cm, 40 cm, 50 cm) and a far range (80 cm). We collect 40 data samples at each range for evaluation, and the results are shown in Fig. \ref{fig: range independence}. At close ranges, both \textit{mmPencil-Qwen} and \textit{mmPencil-LLaVA} maintain high recognition accuracy. Specifically, at 30 cm, the accuracy is 95.00\% for \textit{mmPencil-Qwen} and 90.00\% for \textit{mmPencil-LLaVA}. At 40 cm, the accuracy is 92.50\% and 90.00\%. At 50 cm, \textit{mmPencil-Qwen} achieves 95.00\% accuracy, while \textit{mmPencil-LLaVA} reaches 92.50\%. This is because, at close ranges, the radar can capture the reflection signals of hand movements more clearly, and subtle environmental noise has less impact. Additionally, the multi-stage trajectory reconstruction algorithm refines the spectrum of location data, maintaining high resolution and enabling accurate estimation of handwriting trajectories at different ranges. Notably, even at a distance of 80 cm, both \textit{mmPencil-Qwen} and \textit{mmPencil-LLaVA} maintain an accuracy of 85.00\%, demonstrating the impressive robustness of the \textit{mmPencil} system to variations in range.

\subsubsection{Writing-Angle Independence Evaluation}

To evaluate the system's performance at different writing angles, we also randomly select 8 types of words (including "all," "but," "full," "less," "might," "shown," "almost," and "become") for the experiment. Unlike previous works that target individual characters, word trajectories cover a wider range of angles. To avoid part of the trajectory falling outside the view (60$^\circ$ to the left and right), we choose angles of -30$^\circ$, 0$^\circ$, and 30$^\circ$ for evaluation, as shown in Fig. \ref{fig: angle independence}. At an angle of -30$^\circ$, the accuracy of \textit{mmPencil-Qwen} and \textit{mmPencil-LLaVA} reaches 92.50\% and 97.50\%, respectively. Notably, thanks to the low angle estimation error when facing the radar, both systems achieve 100.00\% accuracy at 0$^\circ$. However, at a 30$^\circ$ angle, the accuracy slightly decreases. This is because handwriting is usually written from left to right, and the last letter may be close to the edge of the view, causing some distortion in the trajectory. Despite this, due to the strong semantic understanding ability of VLMs, the model can effectively correct errors, even if some of the last letters are misrecognized. In this case, the accuracy of \textit{mmPencil-Qwen} and \textit{mmPencil-LLaVA} at 30$^\circ$ is still 87.50\% and 85.00\%, respectively. Overall, the VLM-based \textit{mmPencil} systems show high robustness to angular variations, maintaining high recognition performance even at the field-of-view boundaries.

\subsubsection{Writing-Speed Independence Evaluation}

Writing speed is a key factor that directly affects the smoothness of the trajectory. When the writing speed is high, details of the handwriting trajectory may be lost, making it difficult for the model to accurately reconstruct the trajectory or capture subtle dynamic changes. Therefore, we control the writing time for each character to be 2 seconds, 2.5 seconds, and 3 seconds, and conduct experiments. The results are shown in Fig. \ref{fig: speed independence}. In this set of experiments, we randomly select the following words: "she", "use", "what", "play", "still", "three", "better", and "called". Thanks to the restoration of writing details in trajectory reconstruction and the robustness of VLMs in trajectory feature extraction, both VLMs show balanced performance at different writing speeds. Only \textit{mmPencil-Qwen} shows a slight limitation at higher writing speeds. Specifically, \textit{mmPencil-Qwen}'s accuracy at fast-to-slow writing speeds is 82.50\%, 92.50\%, and 92.50\%, while \textit{mmPencil-LLaVA}'s accuracy is 95.00\%, 92.50\%, and 95.00\%. In practical applications, the writing speed may vary depending on the scenario. Therefore, the system needs to maintain efficient and accurate performance at different speeds. These experiments further demonstrate the robustness of the VLM-based \textit{mmPencil} system under varying writing speeds, laying the foundation for its widespread use in various scenarios.

\begin{figure}
	\centering
        \begin{minipage}[t]{0.315\textwidth}
		\centering
  		\includegraphics[width=\linewidth]
            {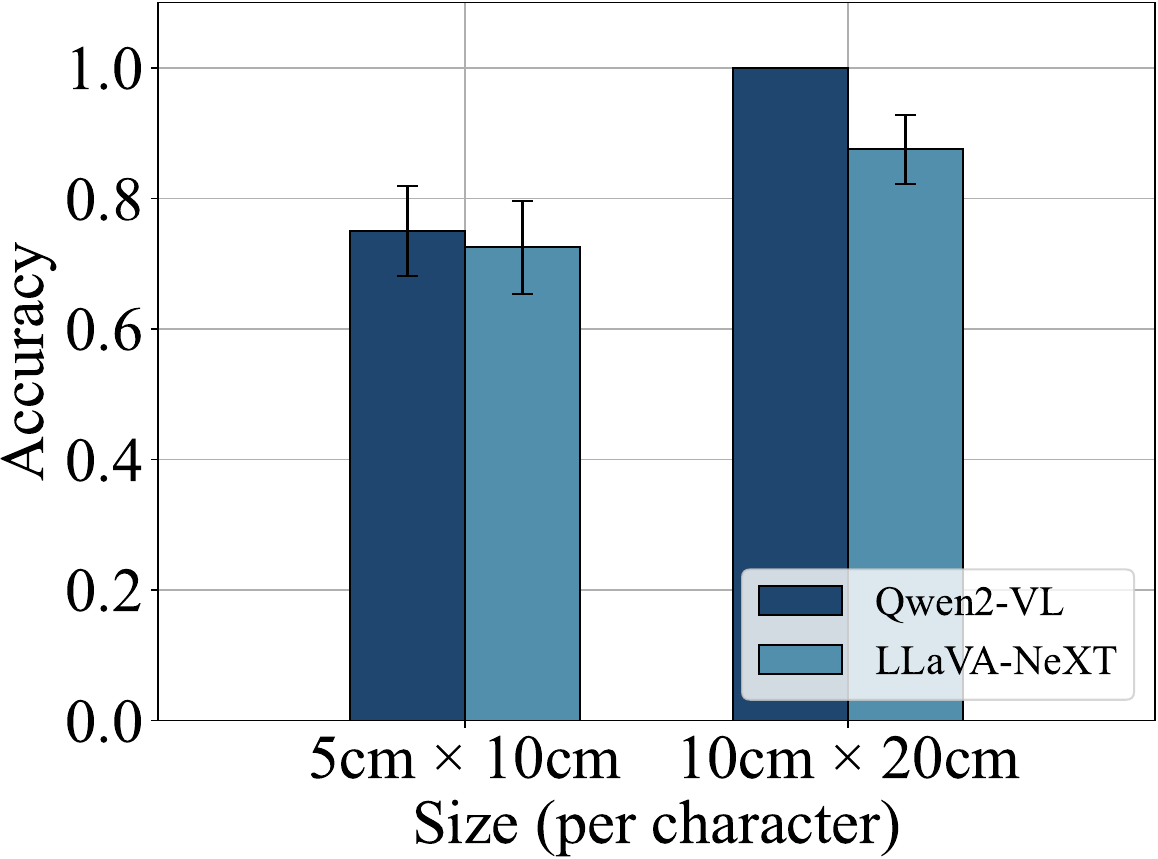}
            \caption{Recognition accuracy  of experiments with different writing sizes.}
  		\label{fig: size independence}
        \end{minipage}\hspace{10mm}
        \begin{minipage}[t]{0.32\textwidth}
		\centering
  		\includegraphics[width=\linewidth]
            {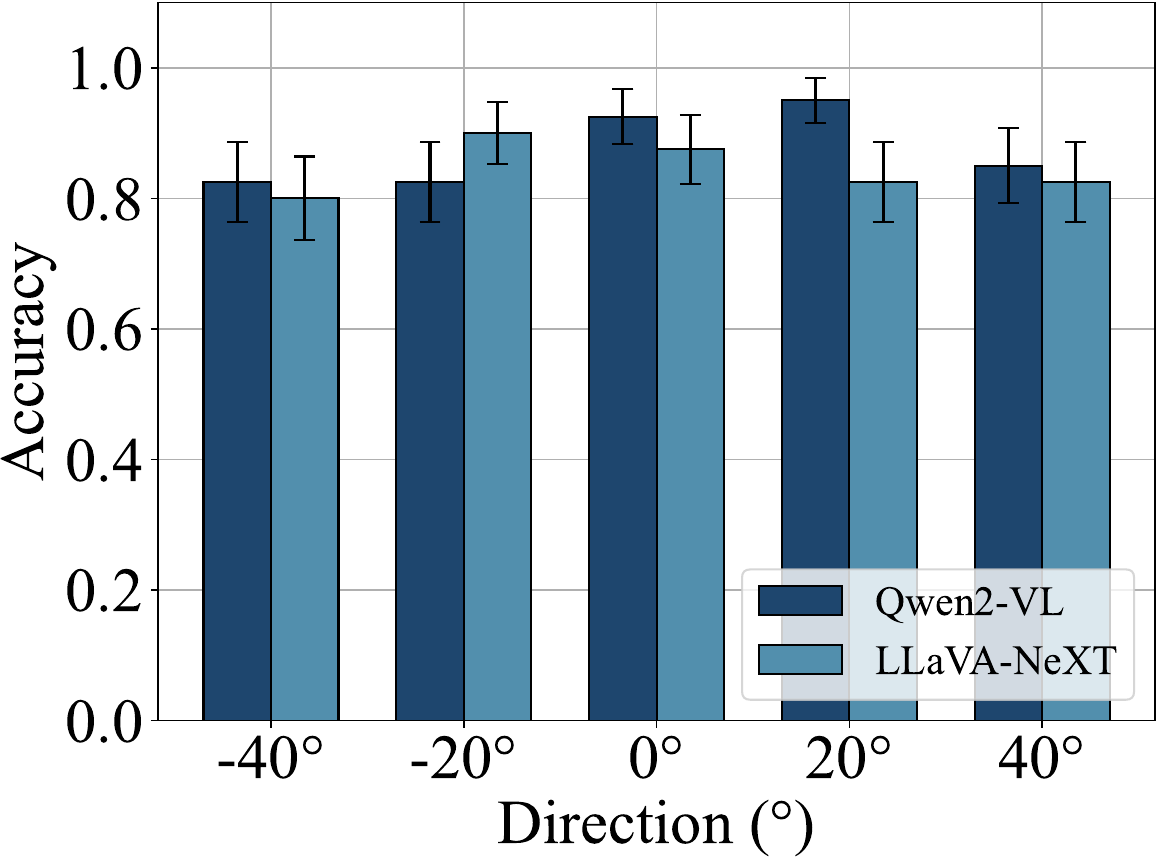}
            \caption{Recognition accuracy  of experiments with different writing directions.}
  		\label{fig: direction independence}
        \end{minipage}
\end{figure}

\subsubsection{Writing-Size Independence Evaluation}

In this section, we consider the effect of writing size on word recognition. Compared to traditional character recognition tasks, word recognition presents greater challenges, especially when the writing size is small, as the connections between letters become more difficult to handle. To evaluate this, we define rectangular writing areas for each letter and test two sizes: 5cm $\times$ 10cm and 10cm $\times$ 20cm. In this set of experiments, we randomly select the following words: "not", "old", "hand", "look", "early", "again", "number", and "strong". The results are shown in Fig. \ref{fig: size independence}. Under normal writing sizes, both \textit{mmPencil-Qwen} and \textit{mmPencil-LLaVA} maintain high recognition accuracies, achieving 100.00\% and 87.50\%, respectively. However, when the writing size is reduced to 5cm $\times$ 10cm, the connecting trajectories between letters overlap with parts of the letters, causing mis-classification. In this scenario, the accuracies drop to 75.00\% for \textit{mmPencil-Qwen} and 72.50\% for \textit{mmPencil-LLaVA}. These results indicate that writing size has a significant impact on trajectory recognition, particularly in extremely small sizes where overlaps and trajectory errors pose challenges to the models. While the system performs well under standard writing sizes, it encounters difficulties in extreme cases, highlighting opportunities for further system optimization in future work.

\subsubsection{Writing-Direction Independence Evaluation}

\begin{figure}
	\centering
	\begin{minipage}[ht]{0.66\textwidth}
		\centering
            \subfigure[A word writing in the direction of -40 $^\circ$ in space.]{
  		\includegraphics[width=0.35\linewidth]{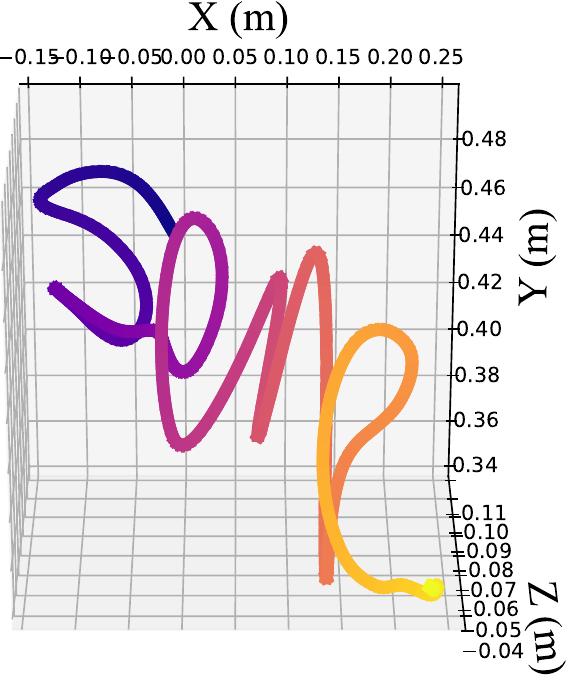}
  		\label{fig: -40}}\hspace{8mm}
            \subfigure[A word writing in the direction of 40 $^\circ$ in space.]{
  		\includegraphics[width=0.35\linewidth]{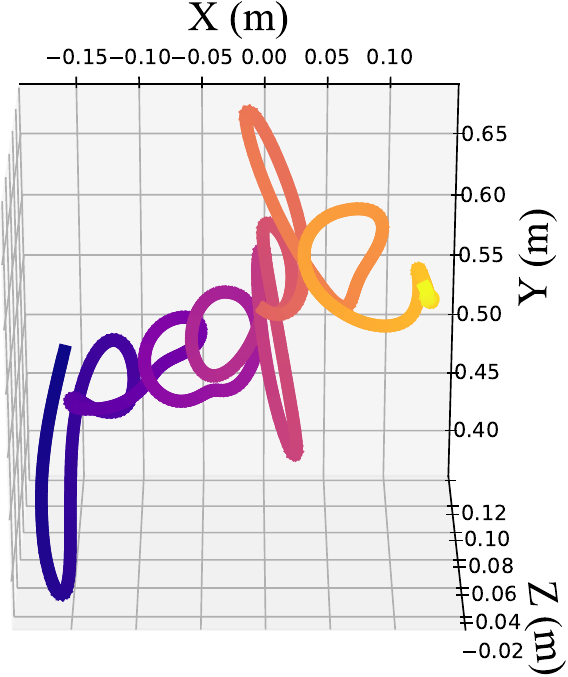}
  		\label{fig: 40}}
            \caption{3D trajectory of words written in direction $\pm$ 40$^\circ$.}
  	\end{minipage}
        \begin{minipage}[ht]{0.32\textwidth}
		\centering
  		\includegraphics[width=\linewidth]
            {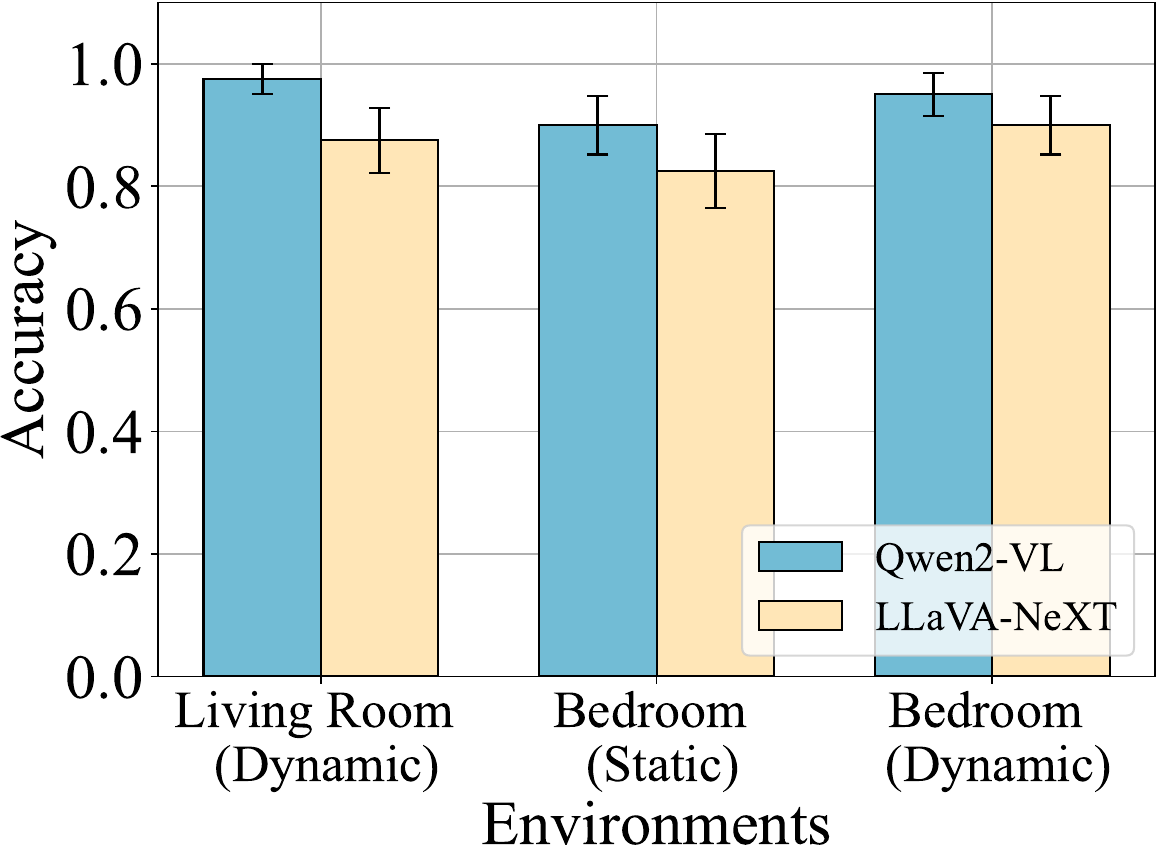}
            \caption{Recognition accuracy at different scenes ("dynamic" and "static" refer to the presence or absence of moving human in the proximity).}
  		\label{fig: environment independence}
        \end{minipage}\hspace{2mm}
\end{figure}

The direction of writing is another challenge in word-level recognition. When writing at an direction, each letter in a word undergoes slight and irregular rotations, as shown in Fig. \ref{fig: -40} and \ref{fig: 40}. This increases the complexity of letter trajectories and causes a certain degree of overlap between letters, which affects the overall recognition performance. It is important to note that data augmentation is not the focus of the \textit{mmPencil} system, so we do not use any data augmentation techniques to address the rotation or similarity issue. Without loss of generality, we select 8 words randomly for experimental evaluation, including "why", "you", "some", "well", "about", "going", "little", and "people". To assess the impact of writing direction, we perform experiments with directions of 0$^\circ$, $\pm$ 20$^\circ$, and $\pm$ 40$^\circ$. The results are shown in Fig. \ref{fig: direction independence}. As the writing direction deviates further, the recognition performance of the systems fluctuates slightly. In particular, the recognition accuracy of \textit{mmPencil-Qwen} swings from 92.50\% at 0$^\circ$ to 95.00\% and 82.50\% at $\pm$ 20$^\circ$, and further to 85.00\% and 82.50\% at $\pm$ 40$^\circ$. Similarly, \textit{mmPencil-LLaVA} shows a comparable trend, achieving an accuracy of 87.50\% at 0$^\circ$, 82.50\% and 90.00\% at $\pm$ 20$^\circ$, and 82.50\% and 80.00\% at $\pm$ 40$^\circ$. This fluctuation is mainly due to the increased difficulty in distinguishing overlapping and rotated trajectories between letters. However, despite the absence of data augmentation, both models demonstrate notable robustness to changes in writing direction. Even at a large deviation of $\pm$ 40$^\circ$, \textit{mmPencil-Qwen} and \textit{mmPencil-LLaVA} maintain recognition accuracies above 80\%. These results highlight the potential and robustness of the \textit{mmPencil} system in handling variations in writing direction, including rotations and tilts. To sum up, due to the strong semantic understanding capabilities of VLMs, the developed system can effectively correct errors caused by mis-recognition of individual letters using the contextual information, thereby preserving the overall performance.

\subsubsection{Environment Independence Evaluation}

The experiments in Sec \ref{performance} are carried out primarily in a living room without human movement in close proximity. To further validate the system’s adaptability to different environments, we perform experiments in various real-life scenes, including a living room scene with a moving individual in proximity and bedroom scenes with and without human movement. Details of the environmental scenes are shown in Fig. \ref{fig: environment}, and the results are shown in Fig. \ref{fig: environment independence}. The results illustrate that \textit{mmPencil-Qwen} achieves recognition accuracies of 97.50\%, 90.00\%, and 95.00\% in the three scenes, respectively. In comparison, the performance of \textit{mmPencil-LLaVA} is slightly worse but still reaches 87.50\%, 82.50\%, and 90.00\%. These outcomes are attributed to the environmental noise cancellation module in the multi-stage trajectory reconstruction algorithm. This module effectively eliminates interference from environmental reflections, ensuring strong robustness across different environments. Accordingly, we can conclude that the \textit{mmPencil} system can be deployed directly in various scenes without requiring prior knowledge or additional training data. This highlights the adaptability and practicality of the system, providing solid support for its application in diverse real-world environments.

\subsubsection{User-Movement Independence Evaluation}
\begin{figure}
	\centering
	\begin{minipage}[t]{0.35\textwidth}
		\centering
  		\includegraphics[width=\linewidth]
            {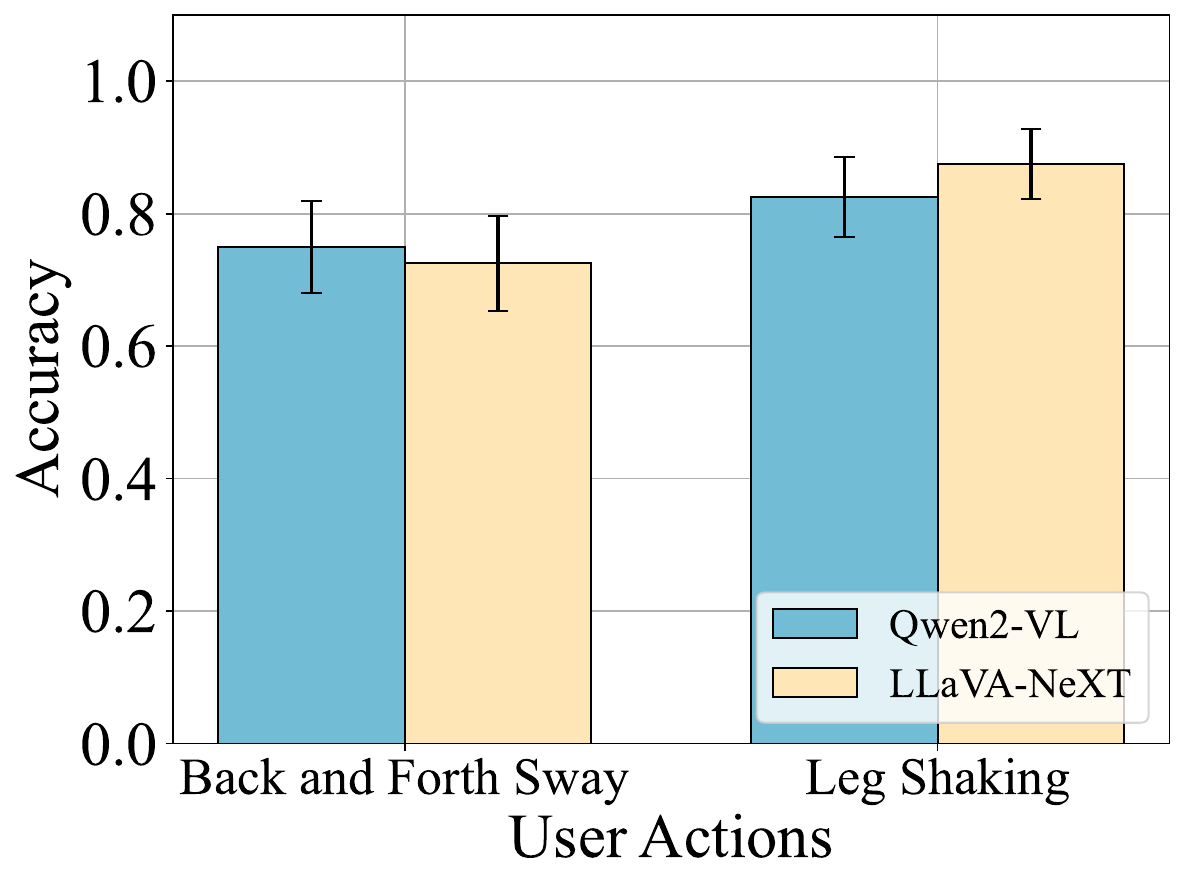}
            \caption{Recognition accuracy  of experiments under different user movements.}
  		\label{fig: user action}
        \end{minipage}\hspace{3mm}
        \begin{minipage}[t]{0.62\textwidth}
		\centering
  		\includegraphics[width=\linewidth]
            {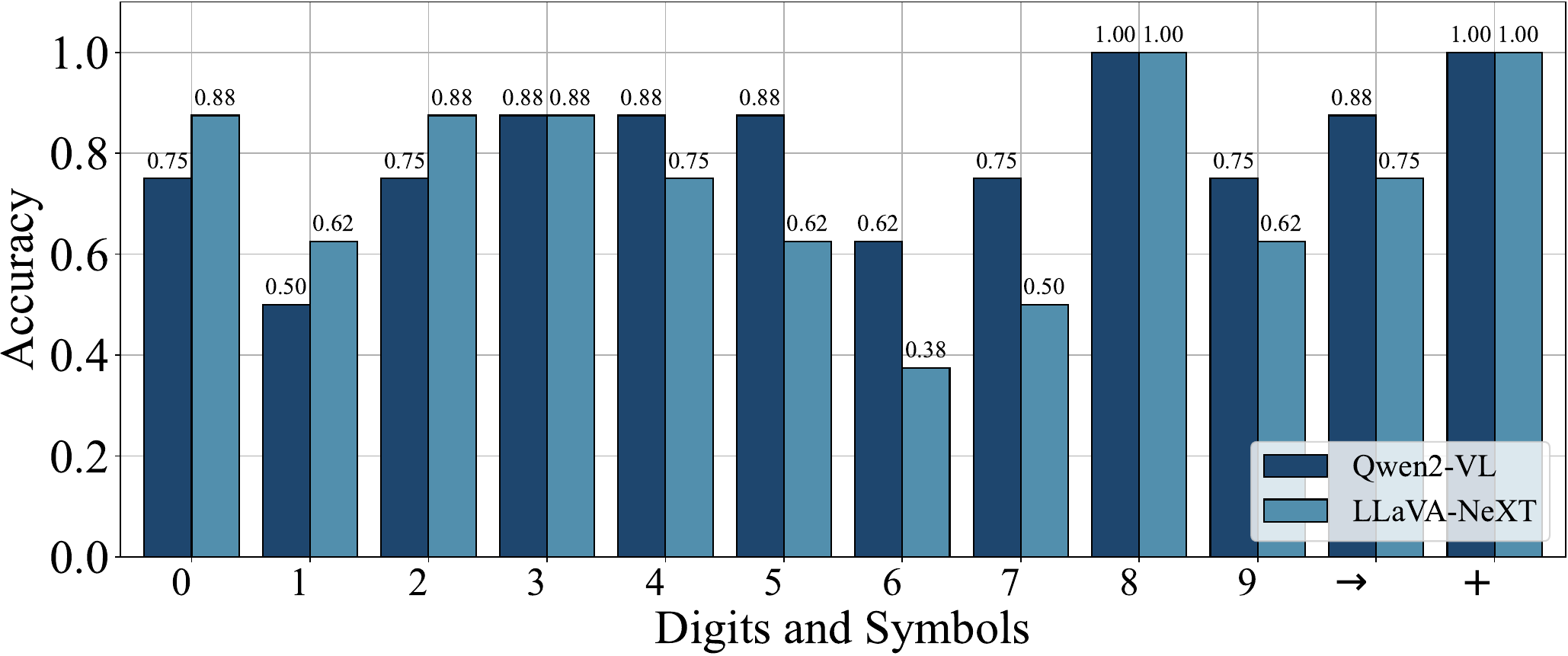}
            \caption{Recognition accuracy of digits and symbols.}
  		\label{fig: number}
        \end{minipage}
\end{figure}

In addition, we evaluate the system performance, in which the user exhibits slight movements while writing words. Specifically, two types of movements are considered, which are body swaying and leg shaking. The corresponding results are shown in Fig. \ref{fig: user action}. We find that body swaying causes significant interference to the system, with the recognition accuracies of \textit{mmPencil-Qwen} and \textit{mmPencil-LLaVA} dropping to 75.00\% and 70.00\%, respectively. In contrast, leg shaking has a relatively smaller impact, resulting in recognition accuracies of 82.50\% and 77.50\%, respectively. This reason is that different types of movements have varying degrees of impact on hand writing actions and upper body swaying causing more pronounced disturbance to the writing trajectory. Overall, slight user movements degrade system performance to a certain extent, and the trajectory smoothing module in the multi-stage spatial trajectory reconstruction algorithm can alleviate part of the fluctuations caused by minor movements. However, the system still struggles to handle scenarios where the user moves significantly, where large trajectory drifts may lead to system failure. To address this issue, we plan to incorporate additional motion information (such as velocity or acceleration signals) to enhance the system's robustness under larger user movements or within mobility scenarios.

\section{Discussion} \label{discussion}

\subsection{Recognition of Handwriting Digits and Symbols}
To further evaluate the capability of \textit{mmPencil} in recognizing simple handwritten elements, we conduct experiments on in-air handwriting digit and symbol recognition. Specifically, we collecte sensing data of the 8 newly 
recruited users covering 10 digits (0–9) and 2 symbols ($\rightarrow$ and $+$), and 4 samples are obtained per digit/symbol per user. As shown in Fig. \ref{fig: number}, the results indicate that the \textit{mmPencil-Qwen} and \textit{mmPencil-LLaVA} models achieve an average recognition accuracy of 80.21\% and 75.00\%, respectively. Generally, the system has promising performance across most categories, where the recognition accuracies of "8" and "$+$" classes are 100\%. However, slightly lower performance is observed for digits with highly similar shapes, such as "1" vs. "7" and "6" vs. "9". Nevertheless, the recognition accuracy for the majority of digits and symbols remains above 75\%, confirming the effectiveness of \textit{mmPencil} in handling low-complexity and few-class recognition tasks. 

It is worth noting that, compared to word recognition, the recognition of digits and symbols inherently presents different challenges. While words typically contain rich contextual information that aids inference and disambiguation, digits and symbols, due to their simpler and often ambiguous stroke patterns, are more prone to confusion. Additionally, the instability and loss of fine-grained details during in-air writing further exacerbate the difficulty. To address these issues, in future work, we plan to introduce structured prior knowledge (e.g., modeling the typical writing order of digits) and fine-grained feature alignment mechanisms, aiming to further enhance the system's performance on digit and symbol recognition.

\subsection{Zero-Shot Recognition of Unseen Words}
\label{sec: unseen words}

\begin{figure}
	\centering
	\begin{minipage}[htbp]{\textwidth}
		\centering
  		\includegraphics[width=\linewidth]{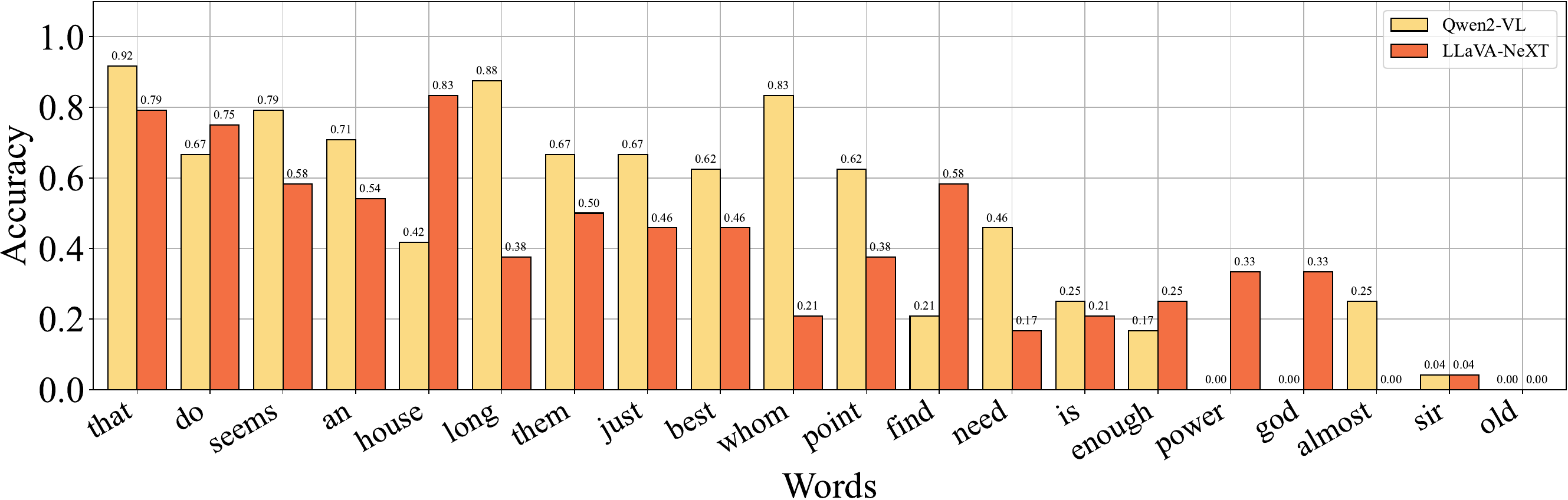}
  		\caption{Zero-shot recognition accuracy of unseen words.}
  		\label{fig: zero shot}
	\end{minipage}
\end{figure}

Given the vast number of words, it is impractical to collect and annotate RF sensing data for every possible word or writing style. Hence, achieving zero-shot recognition of handwriting words becomes essential. Fortunately, the use of VLMs allows for the inference of unseen words by deeply understanding handwriting trajectories and semantic information. Specifically, VLMs excel at aligning visual information with the semantic space of language models through cross-modal learning. As a result, the \textit{mmPencil} system is capable of predicting the structure and meaning of unseen words via semantic associations, facilitating zero-shot recognition.

To evaluate the zero-shot recognition potential of \textit{mmPencil}, we conduct an exploratory experiment. We fine-tune the VLMs using 180 word categories from the \textit{200-Word 4-User Dataset} and evaluate the models with the remaining 20 unseen words. Specifically, we perform instruction fine-tuning on both the \textit{Qwen2-VL} and \textit{LLaVA-NeXT} models. As shown in Fig. \ref{fig: zero shot}, benefiting from the generalisation capability of VLMs, the \textit{mmPencil-Qwen} and \textit{mmPencil-LLaVA} achieve zero-shot word recognition accuracies of 45.83\% and 38.96\%, substantially outperforming the performance of mmPencil-RestNet which almost has no zero-shot recognition capability.

Moreover, we observe significant performance variations across different unseen words. Specifically, words like "that", "do", "seems", and "an" are easily recognized with an accuracy exceeding 70\%, while words like "god", "almost", "sir", and "old" pose challenges for recognition. The reason might be that VLMs tend to prioritize seen words over unseen words during the inference process. For example, if "will" exists in the training data, the model may misclassify the unseen word "well" as "will" due to similarities in spelling and meaning. To address this issue, we plan to redesign the fundamental model architecture and optimize the inference mechanism. The primary focus is on mitigating interference arising from word similarities to enhance zero-shot recognition performance. 

\subsection{Real-Time Processing Feasibility}

\begin{table}[t]
\centering
\caption{Comparison of computational latency and recognition accuracy.}
\label{tab: cost-accuracy}
    \begin{tabular}{ccccc}
    \toprule
    Method & Parameters & \makecell{Trajectory Reconstruction \\ Time (ms per letter)} & \makecell{Inference Time \\ (ms per sample)} & \makecell{Accuracy \\ (\%)} \\
    \midrule
    \textit{ResNet} & 45.5M & 468.30 & 10.05 & 84.13 \\
    \textit{GOT} & 500M & 468.30 & 697.69 & 92.50 \\
    \textit{GOT (moderate SR)} & 500M & 155.65 & 696.73 & 90.17 \\
    \textit{Qwen2.5-VL} & 3B & 468.30 & 33.18 & 83.67 \\
    \textit{Qwen2.5-VL (moderate SR)} & 3B & 155.65 & 32.44 & 80.50 \\
    \textit{Qwen2-VL} & 7B & 468.30 & 185.56 & 89.50 \\
    \bottomrule
    \end{tabular}
    \begin{flushright}
        \footnotesize SR: Spectral Refinement
    \end{flushright}
\end{table}

To evaluate the real-time processing feasibility of the \textit{mmPencil} system in practical applications, we conducted the analysis based on the \textit{50-Word 12-User Dataset} under cross-user evaluation settings, focusing on the trade-offs between computational overhead, inference time, and recognition accuracy, to discuss the practicality of the methods. The overall system latency is primarily attributed to two stages, including the stage of trajectory reconstruction from mmWave signals and the stage of word inference by the classification model on the generated images. 

The experimental results are presented in Table \ref{tab: cost-accuracy}. In the trajectory reconstruction stage, under the standard spectral refinement (SR) strategy, the average reconstruction time per letter is about 468.30 ms. To reduce the latency, we have tested a \textit{moderate SR} strategy, where the degree of spectral refinement is appropriately reduced, decreasing the reconstruction time to 155.65 ms, at the cost of a slight drop in accuracy. During the inference stage, different models exhibit significantly different inference speeds. Specifically, \textit{ResNet-101}, as a lightweight classifier, achieves a fast inference time of only 10.05 ms per sample, with an accuracy of 84.13\%. In contrast, \textit{GOT} takes a longer inference time (697.69 ms per sample) due to the absence of vllm \cite{kwon2023efficient} acceleration, even though it achieves a better recognition performance (92.50\%). It is worth noting that \textit{GOT} has a relatively small parameter size (500M), and once vllm-based acceleration is implemented, its inference time is expected to decrease to a level comparable to \textit{ResNet-101}. For the \textit{Qwen} series models, benefiting from vllm acceleration, Qwen2.5-VL-3B and Qwen2-VL-7B take inference times of 32.44 ms and 185.56 ms per sample, respectively, while maintaining relatively high accuracies of 80.50\% and 89.50\%. To sum up, the \textit{mmPencil} system, configured with \textit{Qwen2.5-VL-3B} and the \textit{moderate SR} strategy, achieves a favorable balance between recognition accuracy and computational latency.

It is important to emphasize that \textit{mmPencil} performs recognition at the word level, with trajectory reconstruction and word typically triggered only after the entire word is written. Based on our experience in data collection, the natural writing time per letter usually ranges between 2 and 3 seconds. Consequently, despite the system's end-to-end latency may exceed 200 ms, it is generally tolerable when compared to the user's natural writing rhythm, ensuring a smooth user experience. We must acknowledge that as we prioritize maximal recognition accuracy, the proposed solution exhibits inherent limitations in real-time processing latency. In future work, we plan to introduce an efficient trajectory reconstruction algorithm based on end-to-end learning and apply model pruning and quantization techniques during inference to further enhance the system's real-time performance, thereby better meeting application scenarios of different latency requirements.

\subsection{In-Air Handwriting Sentence Recognition}
In this study, we have realized word-level handwriting recognition using mmWave radar technology and VLMs. As a result, a logical next step for improvement is advancing towards sentence-level recognition of in-air handwritings. It holds substantial promise for enhancing human-machine interaction, facilitating innovative applications in areas such as education, healthcare, and smart devices. Apparently, sentence-level recognition presents greater complexity and challenges than word-level recognition, due to the intricate semantic structures and contextual information embedded within sentences. Specifically, the challenge mainly lies in extracting meaningful sentence-level features from complex trajectories captured by the mmWave radar. It necessitates efficient segmentation techniques and sophisticated model architectures capable of capturing both local word-level features and global semantic information. Vital to this process are VLMs, which excel at aligning cross-modal data, thereby mapping the captured handwriting trajectories to semantic spaces at the sentence level. Additionally, the acquisition and exploitation of large-scale and diverse RF sensing data through techniques such as data augmentation could play an important role in further enhancing the recognition performance.

\section{Conclusion} \label{conclusion}

In this paper, to achieve writing-style-independent 3D handwriting word recognition, we proposed a system named \textit{mmPencil} based on mmWave radar and VLMs. Specifically, we designed a multi-stage trajectory reconstruction algorithm that fully leverages the rich spatial information of mmWave signals, enabling the extraction of 3D trajectories that are independent of writing position and direction. Additionally, we proposed a spatial-to-visual mapping algorithm to convert the spatial trajectory into 2D visual image that can be processed by VLMs. We collected a mmWave 3D handwriting word dataset and fine-tuned two leading VLMs. Extensive experiments demonstrated that \textit{mmPencil} can achieve accurate word-level handwriting recognition and exhibit high robustness to a wide range of variations, including different users, varying word lengths, diverse writing styles (ranges, angles, speeds, sizes, directions), and environmental conditions and user actions. Finally, we evaluated the real-time processing feasibility of \textit{mmPencil} and explored its potential in zero-shot recognition of digits, symbols, unseen words and sentence-level recognition. 

\begin{acks}
This work is partially supported the National Natural Science Foundation of China (No. 62025205, 62032018, 62472358, 62322601, 62102322) and the Fundamental Research Funds for the Central Universities.
\end{acks}

\bibliographystyle{ACM-Reference-Format}
\bibliography{sample-base}












\end{document}